\documentclass[prd,preprintnumbers,floatfix,aps,nofootinbib,notitlepage,showpacs,twocolumn,superscriptaddress]{revtex4-1}

\usepackage{hhline}
\usepackage[dvipsnames]{xcolor}
\usepackage{amsmath}
\usepackage{hyperref}
\usepackage{graphicx}
\usepackage{bm}
\usepackage{dcolumn}
\newcommand{\bea}{\begin{eqnarray}}
\newcommand{\eea}{\end{eqnarray}}
\newcommand{\be}{\begin{equation}}
\newcommand{\ee}{\end{equation}}

\usepackage{setspace}
\begin{document}

\addtocontents{toc}{\setcounter{tocdepth}{5}} 


\title{Black hole and naked singularity geometries supported by three-form fields}
\author{Bruno J. Barros}
\email{bjbarros@fc.ul.pt}
\affiliation{Instituto de Astrof\'isica e Ci\^encias do Espa\c{c}o, Faculdade de Ci\^encias da Universidade de Lisboa, Edif\'icio C8, Campo Grande, P-1749-016, Lisbon, Portugal}

 \author{Bogdan D\v{a}nil\v{a}}
 \email{bogdan.danila22@gmail.com}
  \affiliation{Astronomical Observatory, 19 Ciresilor Street, 400487 Cluj-Napoca, Romania}
  \affiliation{Department of Physics, Babes-Bolyai University, Kogalniceanu Street,
Cluj-Napoca 400084, Romania,}

 \author{Tiberiu Harko}
 \email{tiberiu.harko@aira.astro.ro}
 \affiliation{Astronomical Observatory, 19 Ciresilor Street, 400487 Cluj-Napoca, Romania}
\affiliation{Department of Physics, Babes-Bolyai University, Kogalniceanu Street,
Cluj-Napoca 400084, Romania,}

\affiliation{School of Physics, Sun Yat-Sen University, Xingang Road, Guangzhou 510275,
P. R. China,}
 \author{Francisco S. N. Lobo}
 \email{fslobo@fc.ul.pt}
 \affiliation{Instituto de Astrofísica e Ci\^encias do Espa\c{c}o, Faculdade de Ci\^encias da Universidade de Lisboa, Edif\'icio C8, Campo Grande, P-1749-016, Lisbon, Portugal}

\date{\today}

\begin{abstract}

We investigate static and spherically symmetric solutions in a gravity theory that extends the standard Hilbert-Einstein action with a Lagrangian constructed from a three-form field $A_{\alpha \beta \gamma}$, which is related to the field strength and a potential term. The field equations are obtained explicitly for a static and spherically symmetric geometry in vacuum. For a vanishing three-form field potential the gravitational field equations can be solved exactly. For arbitrary potentials numerical approaches are adopted in studying the behaviour of the metric functions and of the three-form field. To this effect, the field equations are reformulated in a dimensionless form and are solved numerically by introducing a suitable independent radial coordinate. We detect the formation of a black hole from the presence of a Killing horizon for the time-like Killing vector in the metric tensor components. Several models, corresponding to different functional forms of the three-field potential, namely, the Higgs and exponential type, are considered. In particular, naked singularity solutions are also obtained for the exponential potential case. Finally, the thermodynamic properties of these black hole solutions, such as the horizon temperature, specific heat, entropy and evaporation time due to the Hawking luminosity, are  studied in detail.

\end{abstract}

\maketitle



\section{Introduction}

The use of differential 3-form fields in the realm of cosmology has been gaining more attention over the last decade \cite{Koivisto:2009ew}. These form fields naturally emerge in fundamental theories, such as string theory \cite{Groh:2012tf,Frey:2002qc,Gubser:2000vg} and, thus it is only reasonable to explore their existence under effective formulations of gravity. Its application in cosmology has already proven to be fruitful in explaining the early and late-time acceleration periods of the cosmic history \cite{Koivisto:2009fb,Koivisto:2009sd,DeFelice:2012jt,Germani:2009gg}, reheating \cite{DeFelice:2012wy}, screening solutions \cite{Barreiro:2016aln}, generation of cosmological magnetic fields \cite{Koivisto:2011rm}, among others. Primordial inflation driven by multiple 3-form fields was studied in \cite{Kumar:2014oka}, considering several potential functions. One appealing consequence is that these models present distinct signatures when compared to the standard inflationary setting with a scalar degree of freedom, compatible with recent cosmological observations. Inflationary models in extra dimensional braneworld scenarios inhabited by a single 3-form have also been explored in \cite{Barros:2015evi}, through the use of dynamical systems analysis, and tested against the Planck data. The computation of non-Gaussianities produced by several 3-form fields inflation has been examined in \cite{Kumar:2016tdn} through the analysis of curvature perturbations employing the $\delta N$ formalism.

It is known that in four spacetime dimensions a 3-form field admits a dual scalar field representation \cite{Mulryne:2012ax,Germani:2009iq}. For example, assuming nonquadratic 3-form potentials leads to an equivalent scalar representation exhibiting a noncanonical kinetic term.  However, this mapping is nontrivial, and thus for several self interaction choices, or for any nonminimal coupling, this dual representation breaks down \cite{Koivisto:2009fb}. Nonetheless, even in the cases where the dual scalar description exists, it is often quite complex to deal with and it becomes much more practical and intuitive to work in the form-representative framework.
An interesting feature of 3-forms, is that the standard Maxwell term, $F=d A$, constructed from a massless 3-form $A$, naturally induces a cosmological constant term, and therefore has been used to address the cosmological puzzle regarding the tiny value of $\Lambda$ \cite{Turok:1998he}.
Furthermore, nonminimal interactions between dark energy, driven by a 3-form field, and cold dark matter, was explored in \cite{Koivisto:2012xm,Ngampitipan:2011se} along with the corresponding linear cosmological perturbations.  It was shown that even small values for the coupling lead to substantial variations on the growth of matter fluctuations.

Screening solutions with 3-form fields conformally coupled to matter were also studied in \cite{Barreiro:2016aln}. In \cite{Morais:2016bev} the authors investigate the existence of future abrupt cosmological events, particularly the little sibling of the Big Rip, and the possibility of avoiding these by considering interactions between the 3-form field, portraying dark energy, and dark matter. With the aid of dynamical systems techniques, it was found that this can be achieved only by considering interactions not directly involving the dark matter species. The authors also shed some light on how to distinguish the quadratic and linear dark energy interactions, through the statefinder hierarchy diagnosis and computing the growth of matter perturbations, compatible with the observational SDSS III data. An alternate procedure to avoid these future cosmological abrupt events induced by the 3-form is the quantization of the aforementioned system \cite{Bouhmadi-Lopez:2018lly}. The Lagrangian formalism for a single 3-form fluid with a noncanonical Maxwell term was also explored, in comparison with $k$-essence cosmology, in \cite{Wongjun:2016tva}.

More recently, theories embracing 3-forms have been extended to spherically symmetric and static spacetimes, in particular, wormhole geometries \cite{Barros:2018lca}. More specifically, solutions were found for the modified field equations, assuming a single static and radial-dependent 3-form field, where the standard matter fields are allowed to dwell within the entire wormhole domain without violating the null and weak energy conditions.

Indeed, the present work also deals with static 3-forms supporting spherically symmetric spacetimes, however, in the context of black holes and naked singularities. In fact, black hole solutions are well known in many gravitational field models and, in particular, in standard scalar-tensor extensions of general relativity. For instance, black hole solutions were recently found in the scalar-tensor representation of the hybrid metric-Palatini gravitational theory \cite{Danila:2018xya}, which is a combination of the metric and Palatini $f(R)$ formalisms unifying local constraints at the Solar System level and the late-time cosmic acceleration \cite{Harko:2011nh,Capozziello:2012ny,Capozziello:2015lza,Cambridgebook}. Furthermore, many other exact analytical black hole solutions have been obtained and studied extensively for nonminimally coupled scalar fields  \cite{Fisher:1948yn, Bergmann:1957zza, Janis:1968zz, Bronnikov:1973fh, Solovyev:2012zz, Turimov:2018guy} (for a review of the nonsingular general relativistic solutions with minimally coupled scalar fields see \cite{Bronnikov:2018vbs}).

Generally these latter solutions have been derived in the Einstein frame, without the assumption of the existence of any coupling between the scalar field and the Ricci scalar. Similarly to these scalar field models,  the three-form field theory we have considered is also formulated in the Einstein frame. However,  there are fundamental differences between the three-form field  theory and scalar field models in the Einstein or Jordan frames. An interesting result in the Brans-Dicke type scalar-tensor theories is that the solutions with zero scalar field potential have in general no horizons \cite{Bronnikov:2002kf}. Our analytic and numerical  investigations show that this is not the case in the three-form field  theory. Another interesting result in scalar-tensor theories is that globally regular, asymptotically flat solutions are possible. These solutions correspond to at least partly negative potentials $V(\phi)$, and they are solitons  without horizons and with a regular center \cite{Bronnikov:2002kw}. However, these results specific to scalar-tensor theories cannot be recovered in the three-form fields gravitational theory. On the other hand our analytical and numerical investigations do not indicate the possible existence of any globally regular solutions.

This work is outlined in the following manner: In Sec. \ref{AaF}, we present the general formalism of the three-form field extension of general relativity. In Sec. \ref{back}, considering a static and spherically symmetric background, we deduce the gravitational field equations. In Sec. \ref{secIII:exact}, exact vacuum solutions with three-form fields are presented, for the specific cases of a zero potential and for a constant potential. In Sec. \ref{sec:numerical}, numerical solutions of the field equations are explored, for the Higgs potential and the exponential potential. The thermodynamic properties of the black holes solutions obtained are studied in Sec. \ref{sect5}. Finally,  in Sec. \ref{conclusions}, we discuss and summarize our results.

\section{Einstein gravity with a 3-form field: General formalism}\label{AaF}

We start by considering the following action for Einstein gravity with a standard 3-form field $A_{\alpha\beta\gamma}$,
\begin{equation}
\label{action}
\mathcal{S}=\int d^4 x \sqrt{-g}\left( \frac{1}{2\kappa^2}R+\mathcal{L}_A \right),
\end{equation}
with $R$ being the Ricci scalar, $g=$ det $g_{\mu\nu}$ the determinant of the metric tensor, $\kappa^2 = 8\pi G$ and $\mathcal{L}_A$ stands for the Lagrangian density for our 3-form, which reads \cite{Koivisto:2009fb,Koivisto:2009ew}
\begin{equation}
\label{lagrangian}
\mathcal{L}_A=-\frac{1}{48}F^2 - V(A^2),
\end{equation}
where we have used the notation,
\begin{equation}
F^2 = F_{\alpha\beta\gamma\delta}F^{\alpha\beta\gamma\delta}\quad\quad{\rm and}\quad\quad A^2 = A_{\alpha\beta\gamma}A^{\alpha\beta\gamma}.
\end{equation}
Here $V(A^2)$ is the 3-form potential and  ${\bf F}={\bf dA}$ is the field strength tensor \cite{Morais:2016bev,Mulryne:2012ax}, a 4-form, whose components can be written as
\begin{equation}
\label{maxwell}
 F_{\alpha\beta\gamma\delta} = \nabla_{\alpha}A_{\beta\gamma\delta}-\nabla_{\delta}A_{\alpha\beta\gamma}+\nabla_{\gamma}A_{\delta\alpha\beta}-\nabla_{\beta}A_{\gamma\delta\alpha},
\end{equation}
with $\nabla_{\mu}$ being the covariant derivative.
The equations of motion for our 3-form can be found by varying the action Eq.~\eqref{action} with respect to $A_{\alpha\beta\gamma}$. They read \cite{Koivisto:2009sd,Barros:2018lca}
\begin{equation}
\label{motionT}
\nabla_{\mu} F^{\mu}_{\,\,\,\,\,\,\alpha\beta\gamma}  =12\frac{\partial V}{\partial (A^2)}A_{\alpha\beta\gamma}.
\end{equation}

Varying now Eq.~\eqref{action} with respect to the metric $g^{\mu\nu}$ one finds the following field equations:
\begin{equation}
\label{field}
G_{\mu\nu} =\kappa^2 \, T_{\mu\nu},
\end{equation}
where $G_{\mu\nu}$ is the Einstein tensor and $T_{\mu\nu}$ is the energy momentum tensor of our form field source:
\begin{eqnarray}
T_{\mu\nu} &=& -2 \frac{\delta \mathcal{L}_{A}}{\delta g^{\mu\nu}}+g_{\mu\nu}\mathcal{L}_{A} \nonumber \\
&=& \frac{1}{6}\left( F\circ F \right)_{\mu\nu} + 6 \frac{\partial V}{\partial (A^2)} \left( A\circ A \right)_{\mu\nu} +\mathcal{L}_A\,g_{\mu\nu}, \label{emtensor}
\end{eqnarray}
with a circle denoting contraction of all but the first index, i.e., $\left( F\circ F \right)_{\mu\nu}=F_{\mu\alpha\beta\gamma}F_{\nu}^{\,\,\,\,\alpha\beta\gamma}$ .

We will construct our 3-form with the aid of its dual vector (1-form) \cite{Barros:2018lca}, via the Hodge star operator
\begin{equation}
\label{hodge}
B^{\mu} = (\star A)^{\delta} = \frac{1}{3!}\frac{1}{\sqrt{-g}}\,\epsilon^{\mu\alpha\beta\gamma}A_{\alpha\beta\gamma}.
\end{equation}
Inverting the last identity, we express the three-form components in terms of its dual vector
\begin{equation}
\label{dual}
A_{\alpha\beta\gamma} = \sqrt{-g}\,\epsilon_{\alpha\beta\gamma\delta}B^{\delta}.
\end{equation}
We now follow to build the 3-form components by parameterizing $B^{\mu}$ in terms of a radial scalar function $\zeta = \zeta(r)$, i.e.,
\begin{equation}
\label{duall}
B^{\delta} = \left( 0,\zeta (r),0,0 \right)^{\rm T}.
\end{equation}
Due to the antisymmetric nature of differential forms, once we attain a solution for $\zeta(r)$, all the 3-form components are automatically determined through Eq.~\eqref{dual}.

Through the relations \eqref{maxwell} and \eqref{dual}, we may rewrite the kinetic term in the action Eq.~\eqref{action} as \cite{DeFelice:2012jt}:
\begin{equation}
\label{ah}
-\frac{1}{48}F^2=-\frac{1}{2} F_{0123}F^{0123} = \frac{1}{2}(\nabla_{\mu}B^{\mu})^2.
\end{equation}

We now consider applications of the general formalism obtained here to the specific case of static and spherically symmetric spacetimes.

\section{Spherically symmetric and static background}\label{back}

\subsection{Metric and field equations}

Consider a static and spherically symmetric spacetime, given by the following line element,
\begin{equation}
\label{metric}
ds^2 = -e^{\alpha (r)}dt^2 + e^{\beta (r)}dr^2 + r^2 \left( d\theta^2 + \sin^2\theta \,d\phi^2 \right).
\end{equation}

On this background geometry, the invariant $A^2$ is given by
\begin{equation}
A^2 = -6\, e^{\beta (r)} \zeta (r)^2,
\end{equation}
and the term expressing the kinetic energy of the 3-form is provided by
\begin{equation}
\label{F2}
F^2  = -6\left[ \zeta\left(\alpha' + \beta'  +\frac{4}{r}  \right) + 2\zeta' \right]^2,
\end{equation}
where a prime denotes the derivative with respect to the radial component.

The equations of motion \eqref{motionT} can now be written in terms of $\zeta$, using the metric Eq.~\eqref{metric}, as:
\begin{equation}\label{motion}
2\zeta'' + \left(\alpha'+\beta'+ \frac{4}{r}\right)\zeta'+\left( \alpha''+\beta''-\frac{4}{r^2} \right)\zeta+2V_{,\zeta}=0,
\end{equation}
where $V_{,\zeta} = \partial V / \partial \zeta$.

Through Eq.~\eqref{emtensor}, using Eq.~\eqref{metric}, the components of the energy momentum tensor are then found to be:
\be\label{T1}
T^t_{\,\,\,\,t} = -\rho = \frac{F^2}{48} - V + \zeta V_{,\zeta},
\ee
\be\label{T2}
T^r_{\,\,\,\,r} = p_r = \frac{F^2}{48} - V,
\ee
\be\label{T3}
T^{\theta}_{\,\,\,\,\theta} = T^{\phi}_{\,\,\,\,\phi} = p =  T^t_{\,\,\,\,t},
\ee
where the $F^2$ term is given by Eq.~\eqref{F2} and we identify $\rho$, $p_r$ and $p$ with the energy density, the radial pressure, and the tangential pressure, respectively, of the 3-form. The components of the Einstein tensor can be written as:
\begin{eqnarray}
\hspace{-0.4cm}G^t_{\,\,\,\,t} &=&\frac{1}{r^2}\frac{d}{dr}\left[ r\left( e^{-\beta}-1 \right) \right] = \frac{e^{-\beta}}{r^2}\left( 1-r\beta ' - e^{\beta} \right),  \\
\hspace{-0.4cm}G^r_{\,\,\,\,r} &=& \frac{e^{-\beta}}{r^2} \left(1+r \alpha '-e^{\beta }\right),  \\
\hspace{-0.4cm}G^{\theta}_{\,\,\,\,\theta} &=& G^{\phi}_{\,\,\,\,\phi} = \frac{e^{-\beta}}{2}\left[ \alpha'' + \left( \frac{1}{r}+\frac{\alpha'}{2} \right)\left( \alpha'-\beta' \right) \right] .
\end{eqnarray}

At this point, we have four variables, namely, $\alpha$, $\beta$, $\zeta$ and $V$, and four independent equations, namely, the equation of motion for $\zeta$, i.e., Eq.~\eqref{motion} and the three field equations (setting $\kappa = 1$ for simplicity):
\be
\frac{e^{-\beta}}{r^2}\left( 1-r\beta ' - e^{\beta} \right) =  \frac{F^2}{48} - V + \zeta V_{,\zeta}, \label{fieldI}
\ee
\be
\frac{e^{-\beta}}{r^2} \left(1+r \alpha '-e^{\beta }\right) = \frac{F^2}{48} - V, \label{fieldII}
\ee
\be
\frac{e^{-\beta}}{2}\left[ \alpha'' + \left( \frac{1}{r}+\frac{\alpha'}{2} \right)\left( \alpha'-\beta' \right) \right] = \frac{F^2}{48} - V + \zeta V_{,\zeta}.\label{fieldIII}
\ee

Combining the first two field equations, Eq.~\eqref{fieldI} and Eq.~\eqref{fieldII}, one finds
\begin{equation}\label{sum}
\alpha ' + \beta ' = -re^{\beta}\zeta V_{,\zeta}.
\end{equation}
Using Eq.~\eqref{fieldI} and Eq.~\eqref{fieldIII} yields
\begin{equation}\label{beq}
\frac{2}{r^2}\left(1-r \beta '-e^{\beta}\right) =  \alpha'' + \left( \frac{1}{r}+\frac{\alpha'}{2} \right)\left( \alpha'-\beta' \right).
\end{equation}

Moreover, from Eq.~(\ref{fieldI}) we obtain
\be
r\beta ^{\prime}=1-e^{\beta}\left[1+\left(\frac{F^2}{48}-V+\zeta V_{\zeta}\right)r^2\right].
\ee

These expressions will be useful below.

\subsection{Dynamical system formulation}

The field equation (\ref{fieldI}) can be rewritten as
\be
\frac{d}{dr}\left(re^{-\beta}\right)=1-\left(V-\zeta V_{,\zeta}-\frac{F^2}{48}\right)r^2,
\ee
and can immediately be integrated to give
\be
e^{-\beta}=1-\frac{2M_{{\rm eff}}(r)}{r},
\ee
where the effective mass $M_{{\rm eff}}(r)$ is defined as
\be
M_{{\rm eff}}(r)=\frac{1}{2}\int_0^r{\left(V-\zeta V_{,\zeta}-\frac{F^2}{48}\right)r^2dr},
\ee
and satisfies the mass continuity type equation
\be\label{deq1}
\frac{dM_{{\rm eff}}(r)}{dr}=\frac{1}{2}\left(V-\zeta V_{,\zeta}-\frac{F^2}{48}\right)r^2.
\ee

From Eq.~(\ref{fieldII}) we obtain the expression of $\alpha '$ as
\be\label{deq2}
\alpha '=\frac{r^3\left(F^2/48-V\right)+2M_{{\rm eff}}(r)}{r^2\left(1-2M_{{\rm eff}}(r)/r\right)}.
\ee

With the use of Eq.~(\ref{sum}), Eq.~(\ref{motion}) can be reformulated as
\be\label{deq3}
\zeta ''+\left[\frac{2}{r}-\frac{r\zeta \left(V_{,\zeta}+\zeta V_{,\zeta \zeta}/2\right)}{1-2M_{{\rm eff}}(r)/r}\right]\zeta '-G\left(r,\zeta\right)\zeta+V_{,\zeta}=0,
\ee
where we have denoted
\bea
G\left(r,\zeta \right) &=&\frac{2}{r^{2}}+\frac{\zeta V_{,\zeta }}{%
2\left[1-2M_{{\rm eff}}(r)/r\right]}  \times
	\nonumber\\
&&\Bigg[ 2-\frac{1+\left( F^{2}/48-V+\zeta V_{,\zeta }\right)
r^{2}}{1-2M_{{\rm eff}}(r)/r}\Bigg] .
\eea

Finally, for the function $F^2$ we obtain
\be\label{F}
F^2=-6\left[\left(\frac{4}{r}-\frac{r\zeta V_{,\zeta}}{1-2M_{{\rm eff}}(r)/r}\right)\zeta +2\zeta '\right]^2.
\ee

Here we have presented the relevant equations, which set the stage for exploring solutions, both exactly and numerically.

\section{Exact vacuum solutions with three-form fields}\label{secIII:exact}

In the present Section we will consider some solutions of the system of the vacuum field equations (\ref{fieldI})--(\ref{fieldIII}), respectively, which must be solved together with Eq.~(\ref{motion}), once the functional expression of the three-form potential $V$ is fixed.

\subsection{First case: $V=0$}

If the three-form field potential $V$ identically vanishes, $V\equiv 0$, then  Eq.~(\ref{sum}) can be immediately integrated, giving
\be
\alpha =-\beta,
\ee
where we have set, without loss in generality, the arbitrary integration constant as equal to zero. Then Eq.~(\ref{beq}) becomes
\be\label{28}
\alpha ''+\alpha '^2=\frac{2}{r^2}\left(1-e^{-\alpha}\right).
\ee
By introducing a new variable $\alpha=\ln f$, Eq.~(\ref{28}) becomes
\be\label{f1}
f''-\frac{2f}{r^2}+\frac{2}{r^2}=0,
\ee
with the general solution given by
\be\label{ex1}
f\left(r\right)=1+\frac{c_1}{r}+c_2 r^2,
\ee
where $c_1$ and $c_2$ are arbitrary constants of integration. Since in the limit $F^2\rightarrow 0$ the Schwarzschild solution of standard general relativity must be recovered, it follows that $c_1=-2M$, where $M$ is the mass of the gravitating body, while $c_2=-\Lambda$ can be interpreted as the cosmological constant. Hence, we have recovered the Schwarzschild-de Sitter solution, with the cosmological constant naturally included, as
\be
e^{\alpha}=e^{-\beta}=1-\frac{2M}{r}-\Lambda  r^2.
\ee
Equation~(\ref{motion}), giving the evolution of the function $\zeta$, becomes
\be\label{zetac}
\zeta ''+\frac{2}{r}\zeta ^{\prime}-\frac{2}{r^2}\zeta=0,
\ee
and it has the general solution
\be\label{zeta1}
\zeta (r)= C_1r + \frac{C_2}{r^2},
\ee
where $C_1$ and $C_2$ are arbitrary constants of integration. With the use of Eq.~(\ref{zeta1}),  we obtain finally for $F^2$ the expression
\be\label{ex2}
F^2=-216 C_1^2={\rm constant},
\ee
as expected. On the other hand with the use of the field equation (\ref{fieldII}) we obtain $F^2=144c_1=-288M$.  By comparing the two expressions for $F^2$  we obtain for $C_1$ the representation $C_1=\sqrt{4M/3}$.

\subsection{The constant potential $V=V_0={\rm constant}$}

In the case of the constant potential $V=V_0={\rm constant}$, Eqs.~(\ref{deq3}) take the same form as in the case of the vanishing scalar potential of the three-form field, Eq.~(\ref{zetac}), and its solution is given again by Eq.~(\ref{zeta1}). As for $F^2$, given by Eq.~\ref{F}), we obtain again $F^2=-216C_1^2$. By integrating the mass continuity equation Eq.~(\ref{deq1}), we obtain
\be
M_{{\rm eff}}(r)=\frac{1}{6}\left[V_0+\frac{9}{2}C_1^2\right]r^3+c_1,
\ee
where $c_1$ is an arbitrary constant of integration. Hence for $e^{-\beta}$ we immediately obtain
\be
e^{-\beta}=1-\frac{c_1}{r}-\frac{1}{6}\left[V_0+\frac{9}{2}C_1^2\right]r^2.
\ee

Since also in the constant potential case the general relation $\alpha +\beta=0$ holds, we obtain the metric tensor coefficient $e^{\alpha}$ as
\be
e^{\alpha}=1-\frac{c_1}{r}-\frac{1}{6}\left[V_0+\frac{9}{2}C_1^2\right]r^2.
\ee

The solution is again of the Schwarzschild-de Sitter type, with the potential $V_0$ generating, together with $F^2$, an effective cosmological constant. On the other hand, the arbitrary integration constant $c_1$ is undetermined by the field equations, and must be chosen from physical considerations. If the integration constant $C_1$ and the constant potential $V_0$ vanish, the metric reduces to the standard Schwarzschild form.  Moreover, there are no restrictions on the integration constant $c_1$, whose sign and physical interpretation remains arbitrary.

\section{Numerical solutions of the field equations}\label{sec:numerical}

The system of three equations (\ref{deq1}), (\ref{deq2}) and (\ref{deq3}) for the three unknown functions $M_{{\rm eff}}$, $\alpha $ and $\zeta $, representing a strongly nonlinear system of differential equations, determines the vacuum solutions of the three-form field model gravity. In order to integrate the equations we introduce a new independent variable $\eta $, defined as
\be
\eta =\frac{1}{r}.
\ee

Then we can reformulate the gravitational field equations as the following first order dynamical system
\be\label{d1}
\frac{d\zeta}{d\eta}=u,
\ee
\be\label{d2}
\frac{dM_{{\rm eff}}}{d\eta}=\frac{1}{2}\left(\frac{F^2}{48}+\zeta V_{,\zeta}-V\right)\frac{1}{\eta ^4},
\ee
\be\label{d3}
\frac{d\alpha}{d\eta}=-\frac{F^2/48-V+2\eta ^3M_{{\rm eff}}}{\eta ^3\left(1-2\eta M_{{\rm eff}}\right)},
\ee
\be\label{d4}
\frac{du}{d\eta }=-\frac{V_{,\zeta }+\zeta V_{,\zeta \zeta }/2}{\eta ^{3}\left(
1-2\eta M_{{\rm eff}} \right) }u+\frac{1}{\eta ^{4}}G\left(
\eta ,\zeta \right) \zeta -\frac{V_{,\zeta }}{\eta ^{4}},
\ee
where
\begin{eqnarray}
G\left( \eta ,\zeta \right)& =&2\eta ^{2}+\frac{\zeta V_{,\zeta }}{2\left(1-2\eta
M_{{\rm eff}}\right) }\times \nonumber\\
&&\left[ 2-\frac{\eta ^{2}+\left( F^{2}/48-V+\zeta
V_{,\zeta }\right) }{\eta ^{2}\left( 1-2\eta M_{{\rm eff}}
\right) }\right],
\end{eqnarray}
and
\begin{equation}
F^{2}=-6\left[ \left( 4\eta -\frac{\zeta V_{,\zeta }}{\eta \left( 1-2\eta
M_{{\rm eff}}\right) }\right) \zeta -2\eta ^2u\right] ^{2},
\end{equation}
respectively.  In order to obtain the above equations we have used the mathematical relations $d\zeta/dr=-\eta ^2d\zeta /d\eta$, and
\be
\frac{d^2 \zeta}{dr ^2}=\eta ^4\frac{d^2\zeta }{d\eta ^2}+2\eta ^3\frac{d\zeta}{d\eta},
\ee
respectively. The system of equations (\ref{d1}), (\ref{d2}), (\ref{d3}) and (\ref{d4}) must be integrated with the initial conditions at infinity, given by $M_{{\rm eff}}(0)=M_{{\rm eff}}^{(0)}$, $\alpha (0)=0$, $\zeta (0)=\zeta _0$, and $u(0)=u_0$, respectively.

\subsection{The Higgs potential: $V(\zeta)=\mu ^2\zeta ^2+\nu \zeta ^4$}

\subsubsection{General considerations}

The Higgs-type potential
\be\label{Hpot}
V(\zeta)=\mu ^2\zeta ^2+\nu \zeta ^4,
\ee
plays a fundamental role in elementary particle physics. From a physical point of view we may assume that $-\mu^2 $ represents the mass of the three-form field associated to the gravitational interaction. For the strong interaction case the Higgs self-coupling constant $\nu $ takes the value $\nu \approx 1/8$ \cite{Aad:2015yga}, a value which follows from the analysis of accelerator experiments. But of course in the case of the gravitational models in the presence of a three-form field the values of both $\mu^2$ and $\xi$ may be very different from those suggested by elementary particle physics.

\begin{figure}[htbp]
\centering
\includegraphics[scale=0.7]{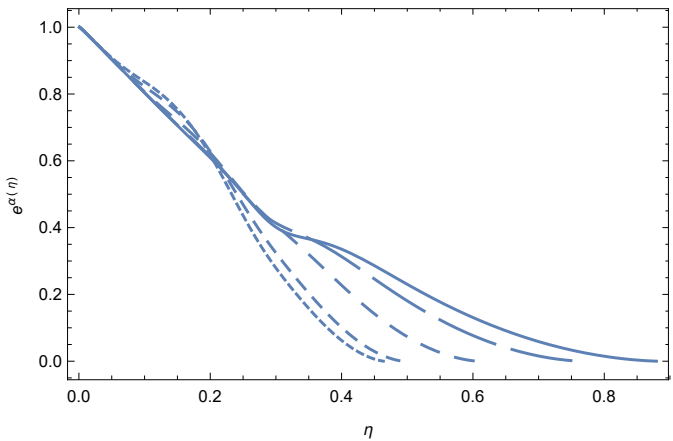}
\caption{Specific case of the Higgs potential: Variation of the metric tensor coefficient $e^{\alpha}$ as a function of the coordinate $\eta$, for $\nu=0.01$ and for different values of $\mu ^2$: $\mu ^2=0.0001$ (solid curve), $\mu ^2=0.00012$ (dotted curve), $\mu ^2=0.00013$ (short dashed curve), $\mu ^2=0.000138$ (dashed curve), and $\mu ^2=0.000141$ (long dashed curve), respectively.
} \label{fig1}
\end{figure}

\begin{figure}[htbp]
\centering
\includegraphics[scale=0.7]{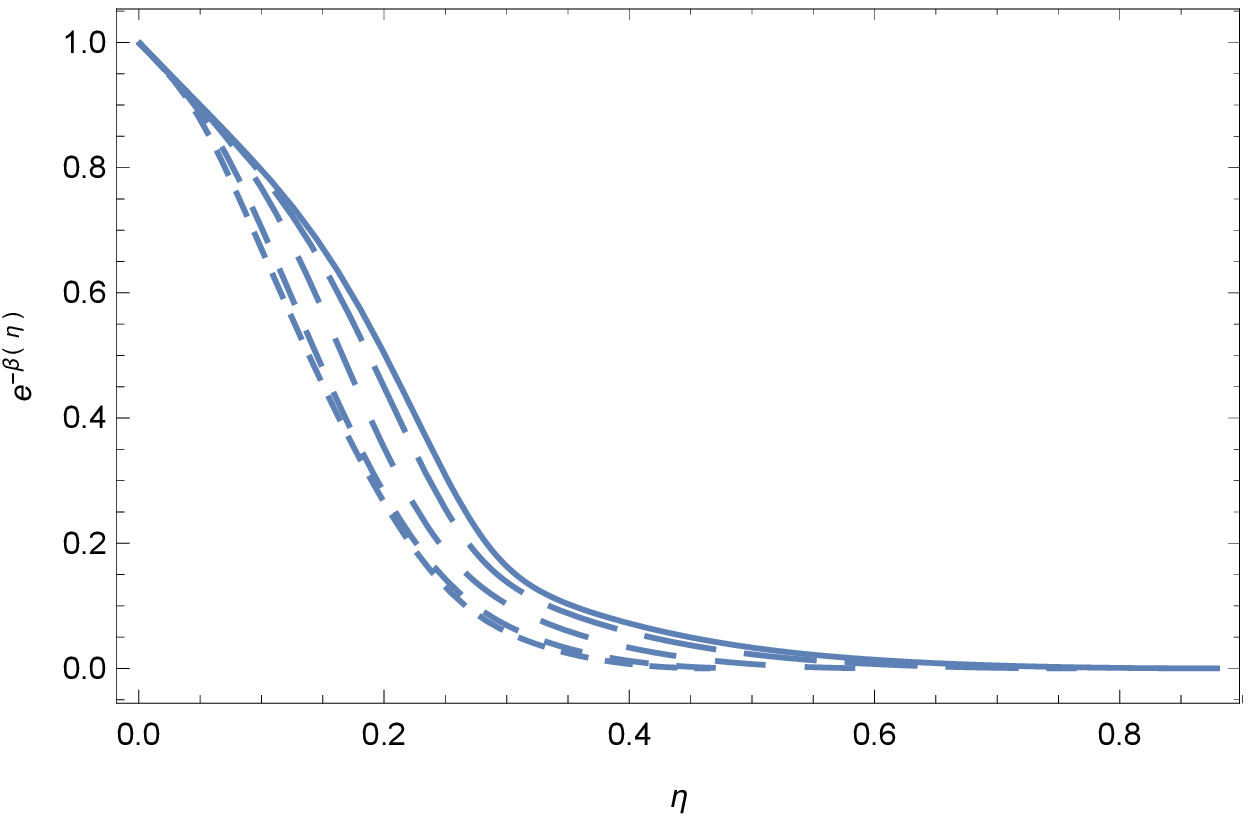}
\caption{Specific case of the Higgs potential: Variation of the metric tensor coefficient $e^{-\beta}$ as a function of the coordinate $\eta$, for $\nu=0.01$, and for different values of $\mu ^2$: $\mu ^2=0.0001$ (solid curve), $\mu ^2=0.00012$ (dotted curve), $\mu ^2=0.00013$ (short dashed curve), $\mu ^2=0.000138$ (dashed curve), and $\mu ^2=0.000141$ (long dashed curve), respectively.
} \label{fig2}
\end{figure}

\begin{figure}[htbp]
\centering
\includegraphics[scale=0.7]{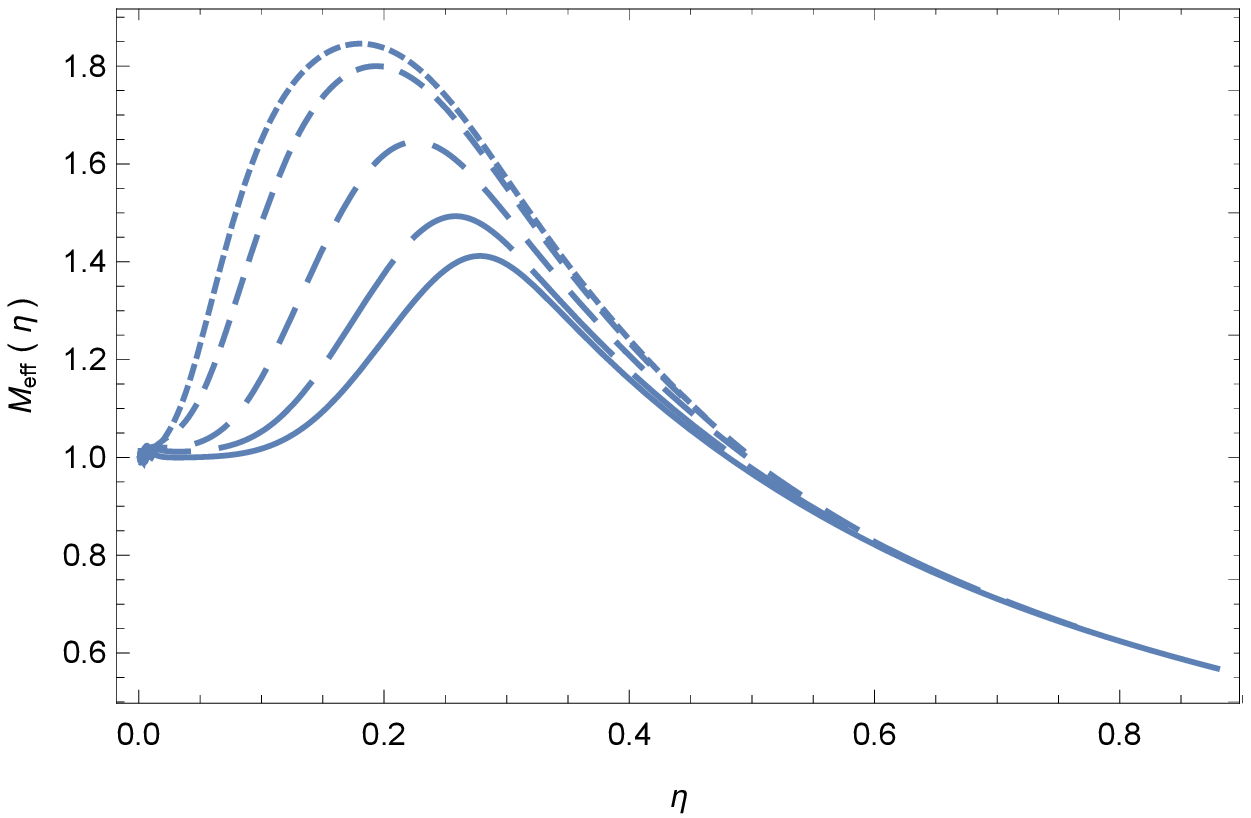}
\caption{Specific case of the Higgs potential: Variation of the effective mass $M_{{\rm eff}}$ as a function of the coordinate $\eta$, for $\nu=0.01$, and for different values of $\mu ^2$: $\mu ^2=0.0001$ (solid curve), $\mu ^2=0.00012$ (dotted curve), $\mu ^2=0.00013$ (short dashed curve), $\mu ^2=0.000138$ (dashed curve), and $\mu ^2=0.000141$ (long dashed curve), respectively.
} \label{fig3}
\end{figure}

\begin{figure}[htbp]
\centering
\includegraphics[scale=0.7]{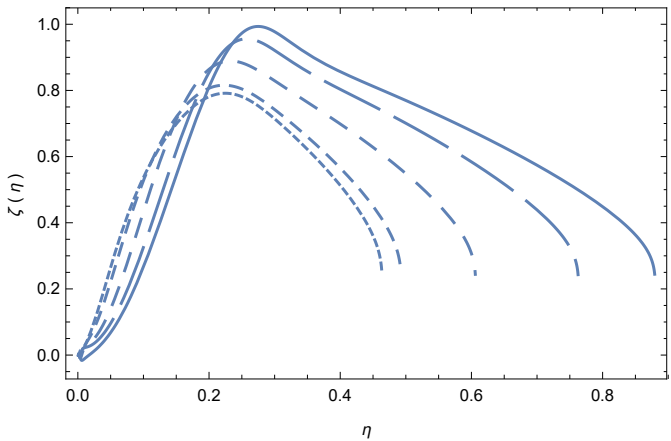}
\caption{Specific case of the Higgs potential: Variation of the radial scalar function  $\zeta$ as a function of the coordinate $\eta$, for $\nu=0.01$, and for different values of $\mu ^2$: $\mu ^2=0.0001$ (solid curve), $\mu ^2=0.00012$ (dotted curve), $\mu ^2=0.00013$ (short dashed curve), $\mu ^2=0.000138$ (dashed curve), and $\mu ^2=0.000141$ (long dashed curve), respectively.
} \label{fig4}
\end{figure}

In the case of the Higgs type potential of the three-form field the vacuum gravitational field equations take the form
\be\label{d1H}
\frac{d\zeta}{d\eta}=u, \quad
\frac{dM_{{\rm eff}}}{d\eta}=\frac{1}{2}\left[\frac{F^2}{48}+\left(\mu ^2+3\nu \zeta ^2\right)\zeta ^2\right]\frac{1}{\eta ^4},
\ee
\be\label{d3H}
\frac{d\alpha}{d\eta}=-\frac{F^2/48-\left(\mu ^2+\nu \zeta ^2\right)\zeta ^2+2\eta ^3M_{{\rm eff}}}{\eta ^3\left(1-2\eta M_{{\rm eff}}\right)},
\ee
\be\label{d4H}
\frac{du}{d\eta }=-\frac{\left(3\mu ^2+10\nu \zeta ^2\right)\zeta}{\eta ^{3}\left(
1-2\eta M_{{\rm eff}} \right) }u+\frac{1}{\eta ^{4}}G\left(
\eta ,\zeta \right) \zeta -\frac{2\left(\mu ^2+2\nu \zeta ^2\right)}{\eta ^{4}},
\ee
where $G(\eta, \zeta)$ and $F^2$ are given by
\begin{eqnarray}
\hspace{-0.5cm}G\left( \eta ,\zeta \right)& =&2\eta ^{2}+\frac{2\left(\mu ^2+2\nu \zeta ^2\right)\zeta ^2}{2\left(1-2\eta
M_{{\rm eff}}\right) }\times \nonumber\\
\hspace{-0.5cm}&&\left\{ 2-\frac{\eta ^{2}+\left[ F^{2}/48+\left(\mu ^2+3\nu \zeta ^2\right)\zeta ^2\right] }{\eta ^{2}\left( 1-2\eta M_{{\rm eff}}
\right) }\right\},
\end{eqnarray}
and
\begin{equation}
F^{2}=-6\left\{ \left[ 4\eta -\frac{2\left(\mu ^2+2\nu \zeta ^2 \right)\zeta ^2}{\eta \left( 1-2\eta
M_{{\rm eff}}\right) }\right] \zeta -2\eta ^2u\right\} ^{2},
\end{equation}
respectively. Eqs.~(\ref{d1H})-(\ref{d4H}) must be considered with the initial conditions at infinity $M_{{\rm eff}}=1$, $\zeta (0)=10^{-5}$, $u(0)=40$, and $\alpha (0)=0$, respectively. In Figs.~\ref{fig1}-\ref{fig4} we present the variations with respect to $\eta$ of the metric tensor coefficients $e^{\alpha}$, $e^{-\beta}$, of the effective mass $M_{{\rm eff}}$ and of the radial scalar function $\zeta$.

The variation of the metric tensor coefficient $e^{\alpha}$ is represented in Fig.~\ref{fig1}. The metric function monotonically decreases from its constant, Minkowskian value at infinity, to zero, a value reached for finite values of $\eta$, and which defines the singular surface of the black hole, or its event horizon. The position of the event horizon is strongly dependent on the numerical values of $\mu ^2$.  A similar behavior characterizes the metric tensor component $e^{-\beta}$, whose variation with respect to $\eta $ is represented in Fig.~\ref{fig2}. The inverse of the metric tensor component decreases linearly from infinity to the event horizon of the black hole, where the metric tensor becomes singular. Similarly to $e^{\alpha}$, the variation of $e^{-\beta}$ is significantly influenced by the numerical values of $\mu ^2$. The changes in the effective mass $M_{{\rm eff}}$ are plotted in Fig.~\ref{fig3}. The mass increases rapidly from its initial value at infinity to a maximum value, reached before the event horizon, an effect due to the presence of the three-form field, and its mass-energy contribution to the mass of the central object. After reaching its maximum value the effective mass decreases before reaching the event horizon. However, for some particular values of $\mu ^2$, the mass becomes approximately constant beginning for some finite value of $\eta$. This indicates that the model enters very quickly in an approximate Schwarzschild regime, with $e^{-\beta}\approx 1-2M_{{\rm eff}}\eta$, with $M_{{\rm eff}}$ a function of the parameters of the Higgs type potential, and of the initial conditions for $\zeta $. This dependence on the initial conditions and on the parameters of the potential also determines the modifications of the position of the event horizon of the black hole, with respect to its standard general relativistic value. The dependence of the scalar radial function $\zeta$ is depicted in Fig.~\ref{fig4}.  The behavior of $\zeta $ indicates a complex dynamics, with $\zeta$ increasing initially, reaching a maximum value, and then becoming again zero at the event horizon. For some values of the Higgs potential the variation has a quasi-oscillatory behavior, characterized by an alternation of local maxima and minima.

\begin{figure}[htbp]
\centering
\includegraphics[scale=0.7]{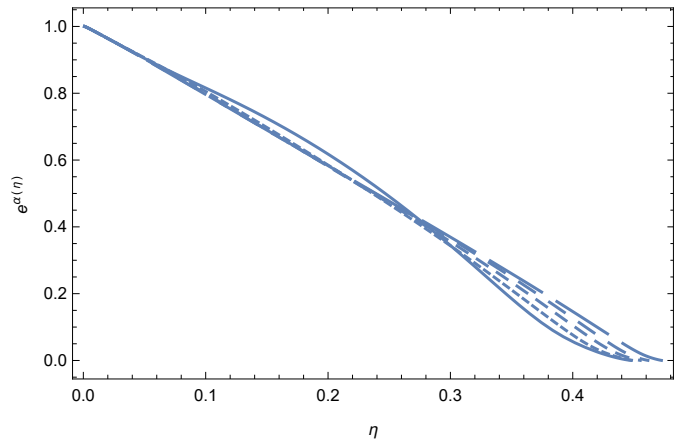}
\caption{Variation of the metric tensor coefficient $e^{\alpha}$, for the Higgs potential, as a function of the coordinate  $\eta$, for $\mu ^2=0.000095$,  and for different values of $\nu $: $\nu=0.04$ (solid curve), $\nu=0.08$ (dotted curve), $\nu=0.12$ (short dashed curve), $\nu=0.16$ (dashed curve), and $\nu (0)=0.20$ (long dashed curve), respectively.
} \label{fig5}
\end{figure}

\begin{figure}[htbp]
\centering
\includegraphics[scale=0.7]{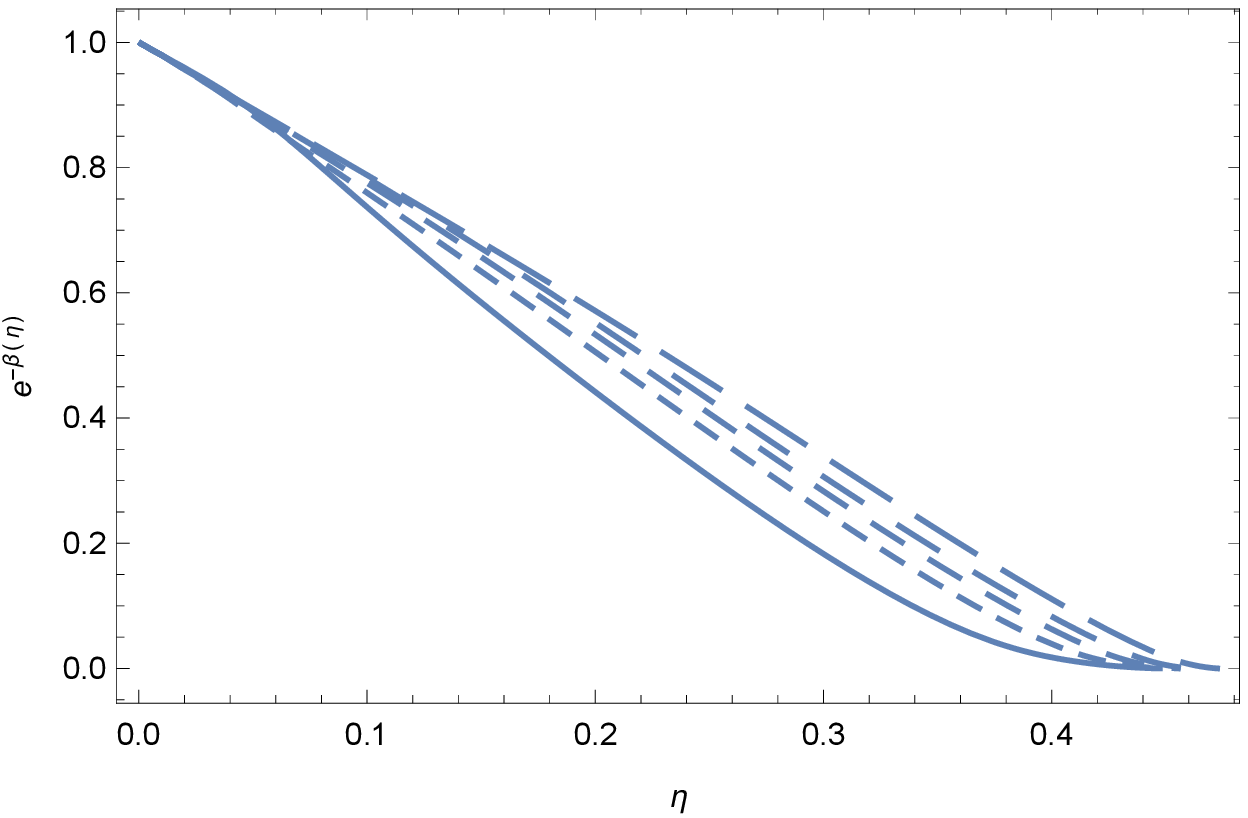}
\caption{Variation of the metric tensor coefficient $e^{-\beta}$, for the Higgs potential, as a function of the coordinate $\eta$,  for $\mu ^2=0.000095$,  and for different values of $\nu $: $\nu=0.04$ (solid curve), $\nu=0.08$ (dotted curve), $\nu=0.12$ (short dashed curve), $\nu=0.16$ (dashed curve), and $\nu (0)=0.20$ (long dashed curve), respectively.
} \label{fig6}
\end{figure}

\begin{figure}[htbp]
\centering
\includegraphics[scale=0.7]{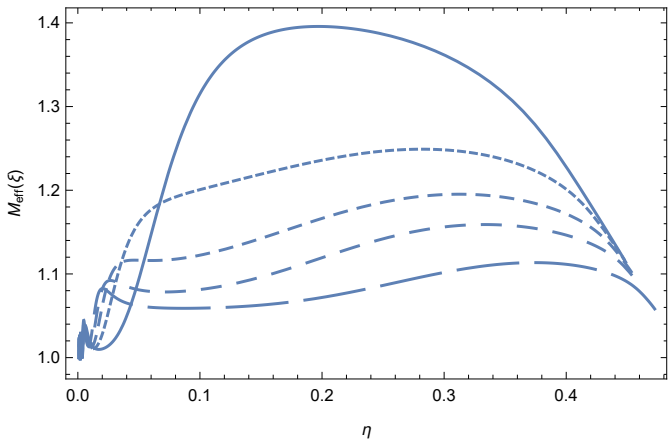}
\caption{Variation of the effective mass $M_{{\rm eff}}$, for the Higgs potential, as a function of the coordinate $\eta$,  for $\mu ^2=0.000095$,  and for different values of $\nu $: $\nu=0.04$ (solid curve), $\nu=0.08$ (dotted curve), $\nu=0.12$ (short dashed curve), $\nu=0.16$ (dashed curve), and $\nu (0)=0.20$ (long dashed curve), respectively.
} \label{fig7}
\end{figure}

\begin{figure}[htbp]
\centering
\includegraphics[scale=0.7]{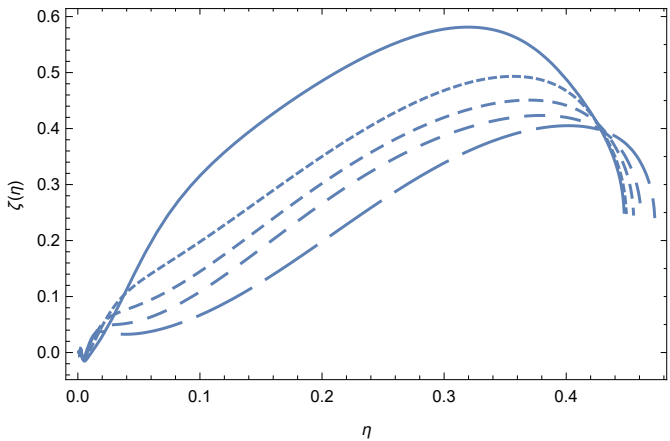}
\caption{Variation of the radial scalar function  $\zeta$, for the Higgs potential, as a function of the  coordinate $\eta$,  for $\mu ^2=0.000095$,  and for different values of $\nu $: $\nu=0.04$ (solid curve), $\nu=0.08$ (dotted curve), $\nu=0.12$ (short dashed curve), $\nu=0.16$ (dashed curve), and $\nu (0)=0.20$ (long dashed curve), respectively.
} \label{fig8}
\end{figure}

The  behavior of the geometric and physical quantities in vacuum for the three-form supported black holes also depend sensitively on the numerical values of the self-coupling constant $\nu$ of the Higgs potential. In Figs.~\ref{fig5}-\ref{fig8} we present the behavior of the metric tensor components, of the effective mass and of the scalar function $\zeta$ for different values of $\nu$.  To integrate the gravitational field equations we have fixed the values of $\mu ^2=0.000095$, $\zeta (0)=0$, $u(0)=40$, $M_{{\rm eff}}(0)=1$, and $\alpha (0)=0$, respectively, and we have varied the values of $\nu$.

The metric tensor components $e^{\alpha}$ and $e^{-\beta}$, represented in Figs.~\ref{fig5} and \ref{fig6}, decrease from their Minkowski values at infinity to zero, corresponding to a finite value of $\eta$, indicating the presence of a singularity corresponding to the formation of a black hole. Their numerical values depend effectively on the numerical values of $\nu$, which also strongly influence the position of the event horizon. The effective mass, shown in Fig.~\ref{fig7}, increases from its value at infinity towards a maximum value reached far away from the event horizon.  The value of the effective mass also strongly depends on the self-coupling constant $\nu$. For some numerical values of $\nu$ the mass becomes roughly a constant beginning from a finite $\eta$,  and the geometry near the compact object becomes quasi-Schwarzschild, with the effective mass of the black hole strongly dependent on the values of $\nu$. The radial scalar function $\zeta$, represented in Fig.~\ref{fig8}, also shows an effective dependence on $\nu$, reaching, similarly to the effective mass, a finite value at the event horizon of the black hole.

\begin{figure}[htbp]
\centering
\includegraphics[scale=0.7]{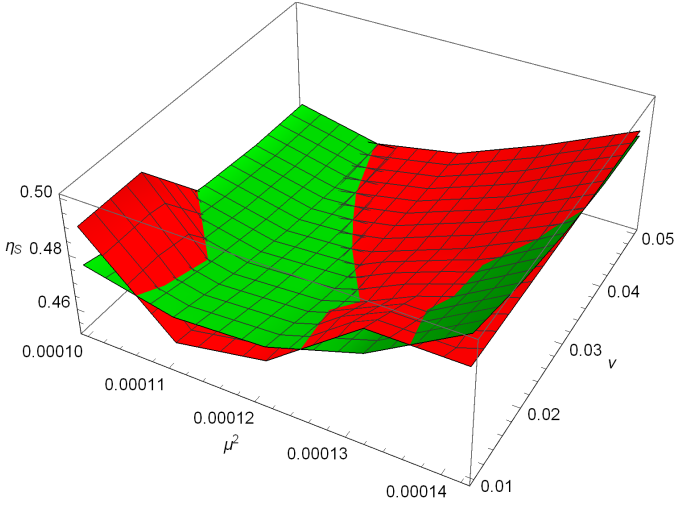}
\caption{Distribution of the event horizons of the three-form field supported black holes as a function of the parameters $\mu ^2$ and $\nu$ of the Higgs potential for $\zeta _0=\{1, 1.105, 1.21, 1.315, 1.42\}\times 10^{-4}$, and $u_0=\{1, 2, 3, 4, 5 \} \times 10^{-2}$.
} \label{fig8a}
\end{figure}

The numerical values of the event horizon $\eta _S$ are presented, for a selected value of the initial conditions and of the model parameters, in Table~\ref{table1}.
In the adopted system of units the position of the Schwarzschild singularity corresponds to $\eta _S=1/2$. The position of the event horizon $r_s$ of the black hole is obtained as $r_s=r_g/\eta _S$, where $r_g$ is the Schwarzschild gravitational radius of the object, defined as $r_g=GM/c^2$, where $M$ is the total mass of the object. For example, the physical position of the event horizon of the black hole supported by a three-form field with Higgs potential having $\eta _S=0.88$ is located at $r_s=1.13r_g$, indicating a black hole more extended that its Schwarzschild counterpart. For a three-form field black hole with $\eta _S=0.44$, the event horizon is located at $r_s=2.27r_g$.
The three-dimensional distribution of the event horizons of the black holes supported by three-form fields is represented in Fig.~\ref{fig8a}.

\begin{table}
\centering
\begin{tabular}{|c | c | c | c | c|}
\hline
$\zeta_0$ & $u_0$ & $\mu^2$ & $\nu$ & $\eta_s$ \\
\hline
$10^{-5}$ & 40  & 0.0001 & 0.01 & 0.88 \\
\hline
$10^{-5}$ & 40  & 0.00012 & 0.01 & 0.46 \\
\hline
$10^{-5}$ & 40  & 0.00013 & 0.01 & 0.49 \\
\hline
$10^{-5}$ & 40  & 0.000138 & 0.01 & 0.61 \\
\hline
$10^{-5}$ & 40  & 0.000141 & 0.01 & 0.76 \\
\hline
$10^{-5}$ & 40  & 0.0001 & 0.015 & 0.59 \\
\hline
$10^{-5}$ & 40  & 0.0001 & 0.02 & 0.51 \\
\hline
$10^{-5}$ & 40  & 0.0001 & 0.015 & 0.48 \\
\hline
$10^{-5}$ & 40  & 0.0001 & 0.015 & 0.46 \\
\hline
$10^{-5}$ & 45  & 0.0001 & 0.01 & 0.75 \\
\hline
$10^{-5}$ & 50  & 0.0001 & 0.01 & 0.66 \\
\hline
$10^{-5}$ & 55  & 0.0001 & 0.01 & 0.58 \\
\hline
$10^{-5}$ & 55  & 0.0001 & 0.01 & 0.54 \\
\hline
$10^{-4}$ & 40  & 0.0001 & 0.01 & 0.78 \\
\hline
$5 \times 10^{-4}$ & 40  & 0.0001 & 0.01 & 0.56 \\
\hline
$10^{-3}$ & 40  & 0.0001 & 0.01 & 0.48 \\
\hline
$5 \times 10^{-5}$ & 40  & 0.0001 & 0.01 & 0.44 \\
\hline
\end{tabular}
\caption{The position of the event horizon $\eta _S$ for selected values of the initial conditions and Higgs model parameters.}\label{table1}
\end{table}

\subsubsection{Interpolating functions}

In order to facilitate the further investigations of the properties of the three-form field supported black holes in the following we will present some explicit analytic expressions for the basic physical and geometrical quantities, obtained from the interpolation of the numerical results. For the effective mass function $M_{{\rm eff}}(\eta)$ we assume a general expression of the form
\begin{equation}
M_{{\rm eff}}(\eta)=A_{MH}+B_{MH}\eta+C_{MH}\eta^2,
\end{equation}
where  the coefficients $A_{MH}$, $B_{MH}$, and $C_{MH}$ are functions of $(\mu^2,\nu,\zeta_0,u_0)$.  In the following analysis we will concentrate mostly on the dependence on the initial conditions, and hence we will fix the parameters of the Higgs potential as $\mu^2=10^{-4}$, and $\nu=10^{-2}$, respectively. Then we obtain
\bea
A_{MH}&=&-1.915 - 3031.75 \zeta_0 - 509.327 \frac{\sqrt{\zeta_0}}{u_0}\nonumber\\
&&+142182 \frac{\zeta_0}{u_0}
+0.1444 u_0-0.0016 u_0^2,
\eea
with the correlation coefficient $R^2=0.999$,
\bea
B_{MH}&=&70.627+81785.6 \zeta_0 - 5.312 \times 10^7 \zeta_0^2 \nonumber\\
&&- 2.98 \times 10^6 \frac{\zeta_0}{u_0} - 3.27 u_0 + 0.036 u_0^2,
\eea
with $R^2=0.989$, and
\bea
C_{MH}&=&-294.508 + 916417 \zeta_0 - 1.4889 \times 10^7 \frac{\zeta_0}{u_0}\nonumber\\
&&+12.86 u_0 - 12456.8 \zeta_0 u_0 - 0.129447 u_0^2,
\eea
with $R^2=0.99$. The general expression of the metric tensor component $e^\alpha$ is
\begin{equation}
e^{\alpha(\eta)}=A_{\alpha H}+B_{\alpha H}\eta+C_{\alpha H}\eta^2,
\end{equation}
with the coefficients $A_{\alpha H}$, $B_{\alpha H}$, $C_{\alpha H}$ given, for fixed $\mu ^2$ and $\nu$,  as functions of $\left(\zeta_0,u_0\right)$  by
\bea
A_{\alpha H}&=&0.847 -125.916 \zeta_0 - 36.981 \frac{\sqrt{\zeta_0}}{u_0}+9656.83 \frac{\zeta_0}{u_0}\nonumber\\
&&+0.0069 u_0-0.000059 u_0^2,
\eea
with $R^2=0.999$,
\bea
\hspace{-0.8cm}B_{\alpha H}&=&0.62+2036.19 \zeta_0 - 4.162 \times 10^6 \zeta_0^2 \nonumber\\
\hspace{-0.8cm}&&- 102912 \times 10^6 \frac{\zeta_0}{u_0} - 0.0942 u_0 + 0.000678 u_0^2,
\eea
with $R^2=0.999$, and
\bea
C_{\alpha H}&=&-2.754 + 90221.2 \zeta_0 - 1.96 \times 10^6 \frac{\zeta_0}{u_0}\nonumber\\
&&+ 0.00058 u_0 - 882.929 \zeta_0 u_0 - 0.00193 u_0^2,
\eea
with $R^2=0.998$.

 For the metric tensor component $e^\beta$ the general interpolating function can be taken as
\begin{equation}
e^{\beta(\eta)}=A_{\beta H}+B_{\beta H}\eta+C_{\beta H}\eta^2,
\end{equation}
with the coefficients $A_{\beta H}$, $B_{\beta H}$, and $C_{\beta H}$ given as functions of $\left(\zeta_0,u_0\right)$ by
\bea
A_{\beta H}&=&1.653 + 730.135 \zeta_0 + 67.886 \frac{\sqrt{\zeta_0}}{u_0}- 35382 \frac{\zeta_0}{u_0} \nonumber\\
&&+0.0324 u_0-0.000353 u_0^2,
\eea
with $R^2=0.999$,
\bea
B_{\beta H}&=&12.688 - 16943.1 \zeta_0 + 9.355 \times 10^6 \zeta_0^2 + 737250 \frac{\zeta_0}{u_0} \nonumber\\
&&- 1.12 u_0 + 0.0296 u_0^2 - 0.000243 u_0^3,
\eea
with $R^2=0.976$, and
\bea
C_{\beta H}&=&59.249 + 212369 \zeta_0 + 3.0549 \times 10^6 \frac{\zeta_0}{u_0} - 2.506 u_0 \nonumber\\
&&+ 2762.06 \zeta_0 u_0 + 0.0206 u_0^2,
\eea
with $R^2=0.999$.

\subsection{The exponential potential: $V(\zeta)=V_0e^{\lambda \zeta}$}

As a second example of black hole solutions supported by a three-form field with non-zero potential, we consider the case of the exponential type potential, $V (\zeta) = V_0e^{\lambda \zeta}$,  where $V_0$ and $\lambda$ are constants. There are many physical processes in string theory and elementary particle physics described by this type of potential. For example, an exponential type potential is obtained in string type theories and in four-dimensional effective Kaluza-Klein theories from the compactification of the higher dimensions.  Moduli fields and non-perturbative effects in quantum field theory such as gaugino condensation can also generate exponential type potentials for scalar fields \cite{deCarlos:1992kox}.  The role of the exponential potential has been intensively investigated especially in the framework of scalar field cosmological and gravitational models and for many field configurations,  including the  inhomogeneous and homogeneous scalar fields \cite{Chen:2000gaa,Rubano:2003et,Gorini:2003wa,Andrianov:2011fg,Andrianov:2012az,Harko:2013gha,Rebesh:2019pbw,Joseph:2019icj,Harko:2020oxq}. In the presence of an exponential potential the gravitational field equations (\ref{d1})-(\ref{d4}) take the form
\be\label{de1}
\frac{d\zeta}{d\eta}=u, \quad  \frac{dM_{{\rm eff}}}{d\eta}=\frac{1}{2}\left[\frac{F^2}{48}+V_0\left(\lambda \zeta -1\right)e^{\lambda \zeta}\right]\frac{1}{\eta ^4},
\ee
\be\label{de2}
\frac{d\alpha}{d\eta}=-\frac{F^2/48-V_0e^{\lambda \zeta}+2\eta ^3M_{{\rm eff}}}{\eta ^3\left(1-2\eta M_{{\rm eff}}\right)},
\ee
\be\label{de3}
\frac{du}{d\eta }=-\frac{V_0\lambda\left(1+\lambda \zeta/2\right) e^{\lambda \zeta} }{\eta ^{3}\left(
1-2\eta M_{{\rm eff}} \right) }u+\frac{1}{\eta ^{4}}G\left(
\eta ,\zeta \right) \zeta -\frac{\lambda V_0e^{\lambda \zeta}}{\eta ^{4}},
\ee
where
\begin{eqnarray}
\hspace{-0.5cm}G\left( \eta ,\zeta \right)& =&2\eta ^{2}+\frac{\lambda V_0\zeta e^{\lambda \zeta}}{2\left(1-2\eta
M_{{\rm eff}}\right) }\times \nonumber\\
\hspace{-0.5cm}&&\left\{ 2-\frac{\eta ^{2}+\left[ F^{2}/48+V_0\left(\lambda \zeta -1\right)e^{\lambda \zeta}\right] }{\eta ^{2}\left( 1-2\eta M_{{\rm eff}}
\right) }\right\},
\end{eqnarray}
and
\begin{equation}
F^{2}=-6\left\{ \left[ 4\eta -\frac{\lambda V_0\zeta e^{\lambda \zeta}}{\eta \left[ 1-2\eta
M_{{\rm eff}}\right] }\right] \zeta -2\eta ^2u\right\} ^{2},
\end{equation}
respectively.

\subsubsection{Naked singularity solutions}

The description of the state and structure of ordinary material systems, forming an initial regular distribution,  after the gravitational collapse, is one of the most important theoretical and observational problems in general relativity. There are two questions one should consider when investigating the gravitational collapse. The first question is to find out the initial conditions of the gravitational collapse that lead to the formation of a black hole. On the other hand a careful investigation of the gravitational collapse shows that it does not end always with the creation of a black hole.  Depending on the initial conditions, another type of object, called a naked singularity,  can also be born as the final state of the collapse \cite{Christodoulou:1984mz, Ori:1987hg,Choptuik:1992jv,Husain:1995bf,Harko:2000ni,Harko:2013sea}. For reviews of the naked singularity problem see \cite{Joshi:2008zz} and \cite{Joshi:2012mk}, respectively.

 The second question one must also necessarily consider is the question if the physically realistic collapse solutions of the Einstein gravitational field  equations that indicate the formation of naked singularities do really correspond to existing natural objects, which can be observed by astrophysical or astronomical methods. If detected observationally, the existence of the naked singularities would be counterexamples of the Cosmic Censorship Hypothesis, proposed by Roger Penrose \cite{Penrose:1969pc}. The Cosmic Censorship Hypothesis  conjectures that curvature singularities are always covered in asymptotically flat spacetimes by event horizons. In fact, one can formulate the Cosmic Censorship Hypothesis in a strong sense (in a geometry that is  physically appropriate naked singularities cannot form), and in a weak sense (if naked singularities do really exist, they are securely covered by an event horizon, and therefore they cannot be detected by far-away observers). There have been many attempts to prove the Cosmic Censorship Hypothesis  (see \cite{Joshi:1987wg} for a review of the early investigations and results in this field). For the possibilities of observationally identifying naked singularities see \cite{Shahidi:2020bla}, and references therein.

\begin{figure}[htbp]
\centering
\includegraphics[scale=0.7]{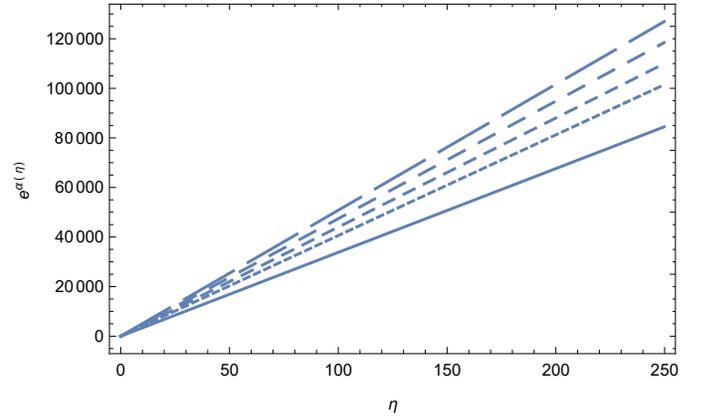}
\caption{Variation of the metric tensor coefficient $e^{\alpha}$ as a function of the coordinate  $\eta$ for the case of the exponential three-form potential $V=V_0e^{\lambda \zeta}$, for $\lambda =-10^{-3}$,  and for different values of $V_0 $: $V_0=0.001$ (solid curve), $V_0=0.0012$ (dotted curve), $V_0=0.0013$ (short dashed curve), $V_0=0.0014$ (dashed curve), and $V_0=0.0015$ (long dashed curve), respectively.
} \label{fig9}
\end{figure}

\begin{figure}[htbp]
\centering
\includegraphics[scale=0.7]{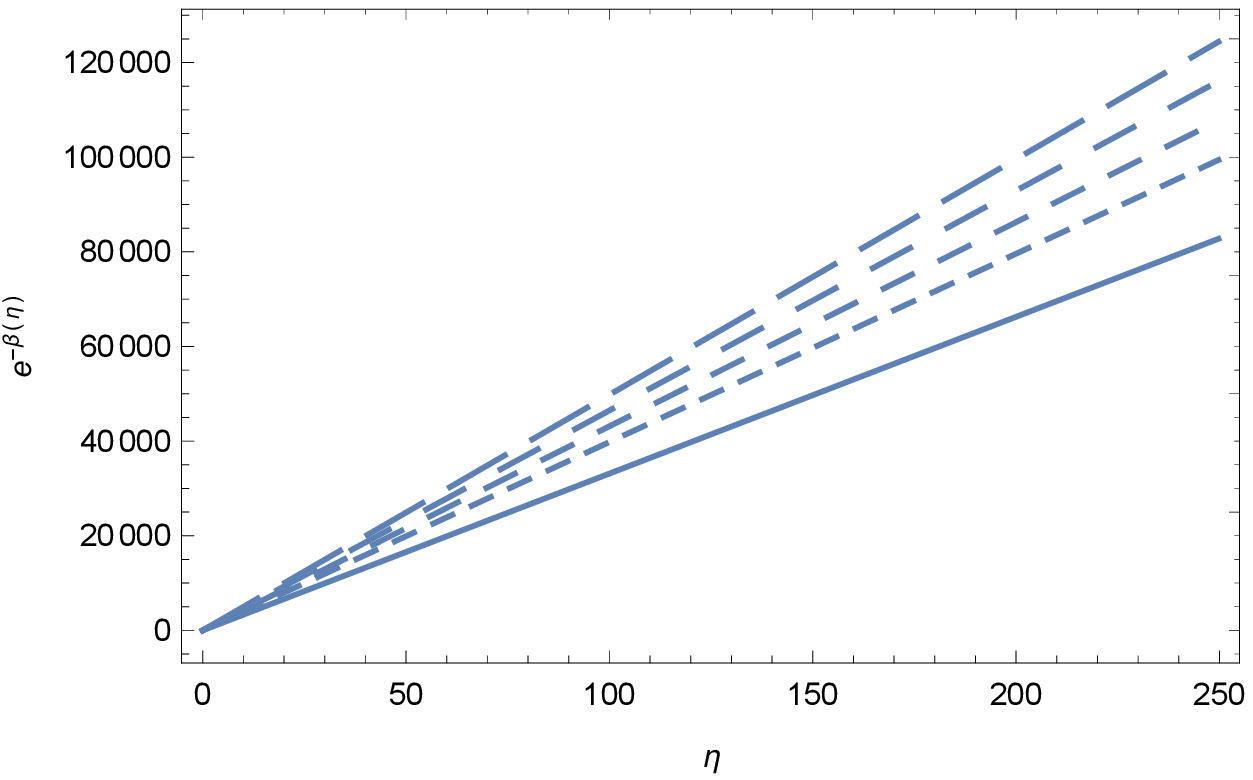}
\caption{Variation of the metric tensor coefficient $e^{-\beta}$ as a function of the coordinate $\eta$ for the case of the exponential three-form potential $V=V_0e^{\lambda \zeta}$, for $\lambda =-10^{-3}$,  and for different values of $V_0 $: $V_0=0.001$ (solid curve), $V_0=0.0012$ (dotted curve), $V_0=0.0013$ (short dashed curve), $V_0=0.0014$ (dashed curve), and $V_0=0.0015$ (long dashed curve), respectively.
} \label{fig10}
\end{figure}

\begin{figure}[htbp]
\centering
\includegraphics[scale=0.7]{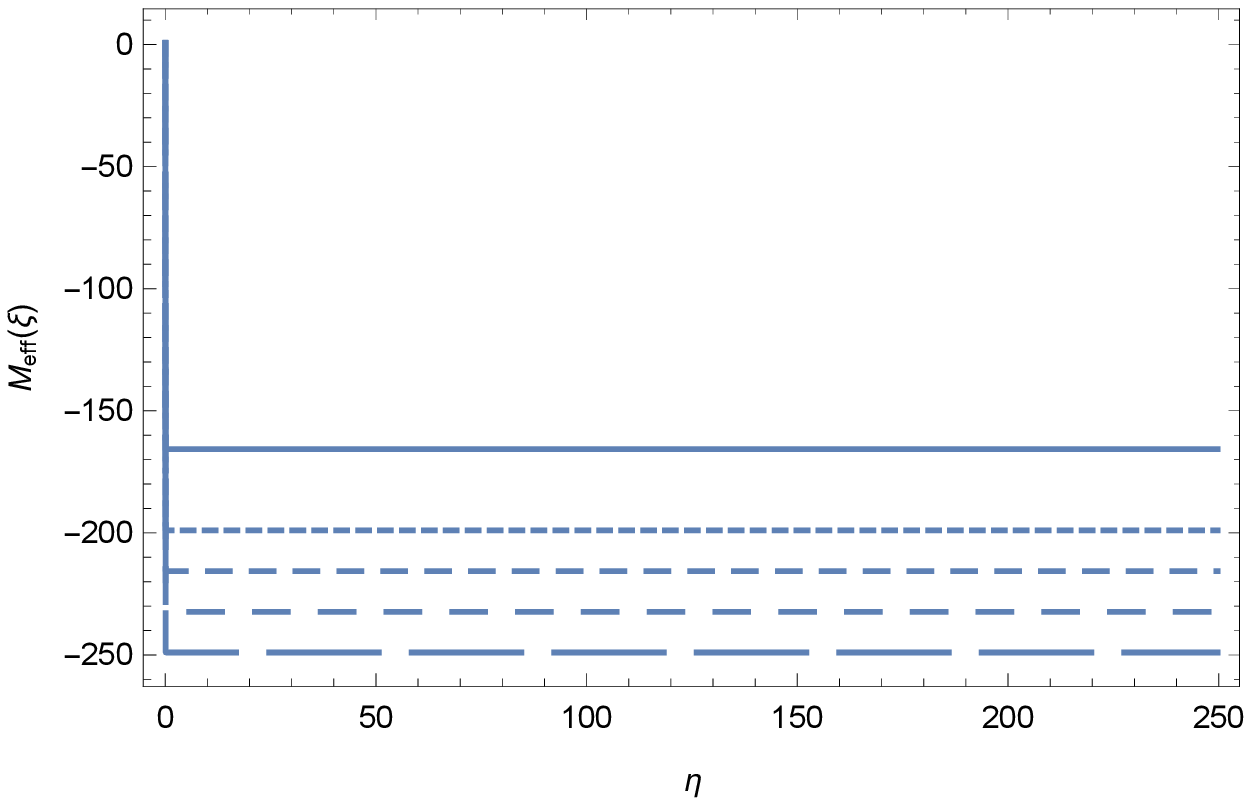}
\caption{Variation of the effective mass $M_{{\rm eff}}$ as a function of the coordinate $\eta$ for the case of the exponential three-form potential $V=V_0e^{\lambda \zeta}$, for $\lambda =-10^{-3}$,  and for different values of $V_0 $: $V_0=0.001$ (solid curve), $V_0=0.0012$ (dotted curve), $V_0=0.0013$ (short dashed curve), $V_0=0.0014$ (dashed curve), and $V_0=0.0015$ (long dashed curve), respectively.
} \label{fig11}
\end{figure}

\begin{figure}[htbp]
\centering
\includegraphics[scale=0.7]{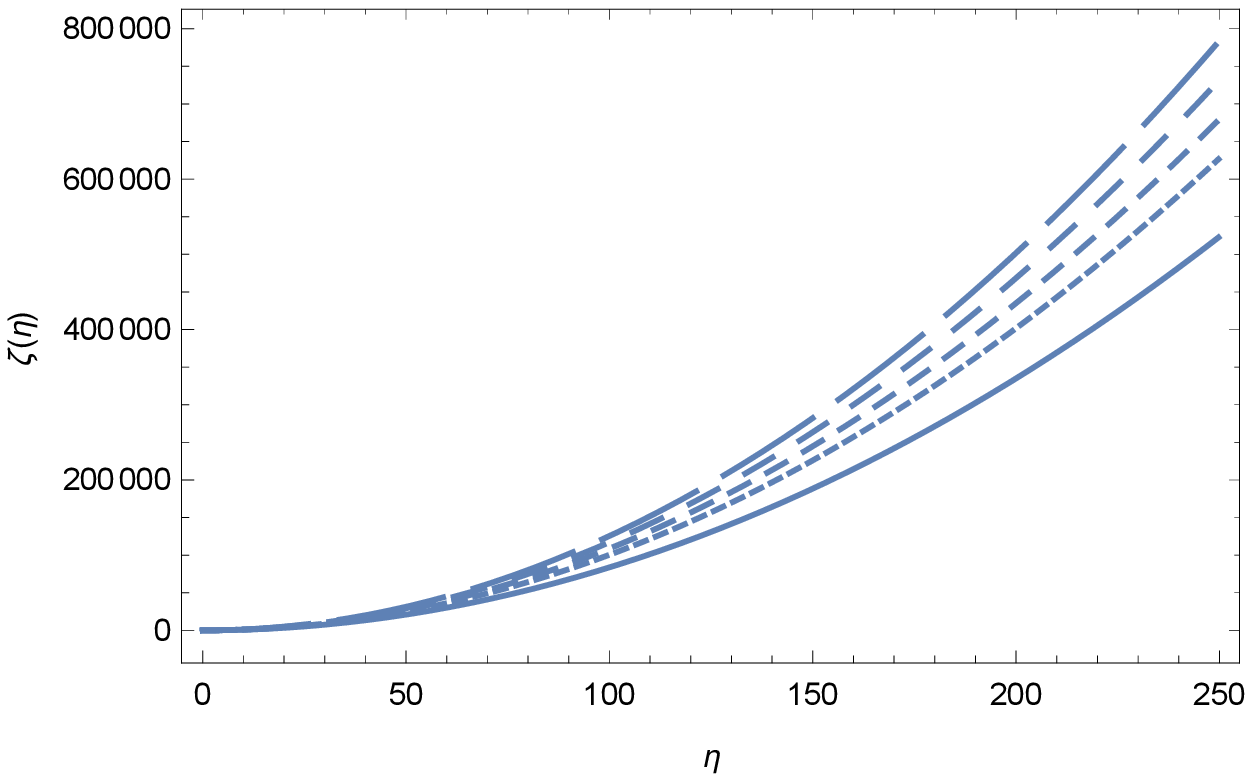}
\caption{Variation of the radial scalar function  $\zeta$ as a function of the  coordinate $\eta$ for the case of the exponential three-form potential $V=V_0e^{\lambda \zeta}$, for $\lambda =-10^{-3}$,  and for different values of $V_0 $: $V_0=0.001$ (solid curve), $V_0=0.0012$ (dotted curve), $V_0=0.0013$ (short dashed curve), $V_0=0.0014$ (dashed curve), and $V_0=0.0015$ (long dashed curve), respectively.
} \label{fig12}
\end{figure}

 For a certain range of parameters and of initial conditions, naked singularity solutions of the gravitational field equations in the presence of a three-form field with exponential potential  can also be obtained. In Figs.~\ref{fig9}-\ref{fig12} we present the behavior of the metric tensor coefficients $e^{\alpha}$, $e^{-\beta}$, of the effective mass $M_{{\rm eff}}$, and of the radial scalar function $\zeta$ for the initial conditions $M(0)=1$, $\alpha (0)=0$, $\zeta (0)=10^{-5}$, and $u(0)=10^{-4}$, respectively. For $\lambda$ we have adopted the value $\lambda =-10^{-3}$, and we have slightly varied the numerical values of $V_0$.

As one can see from Figs.~\ref{fig9} and \ref{fig10}, the metric tensor coefficients are monotonically increasing functions of $\eta$, and they are singular only at the origin $\eta \rightarrow \infty$, or, equivalently, $r\rightarrow 0$. The correspondent massive object does not have an event horizon, and therefore it corresponds to a naked singularity, with the only singular point located at the center. The effective mass of the naked singularity, presented in Fig.~\ref{fig11}, becomes negative at infinity, and takes a constant, negative value up to the singular center of the naked singularity. Hence the metric of this exotic object can be represented as
\be
e^{\alpha (r)}=e^{-\beta (r)}=1+\frac{2M_{{\rm eff}}\left(V_0,\lambda, \zeta _0,\zeta '(0)\right)}{r}.
\ee
The numerical values of the (negative) effective mass are determined by the initial conditions of the gravitational field equations at infinity, as well as by the parameters of the exponential potential. The solutions of the gravitational field equations depend sensitively on these parameters. The radial scalar function $\zeta$, shown in Fig.~\ref{fig12}, diverges at the center of the naked singularity. Its behavior is also dependent on the initial conditions used to solve the gravitational field equations, and on the parameters of the exponential potential.

\subsubsection{Black hole solutions}

\begin{figure}[htbp]
\centering
\includegraphics[scale=0.7]{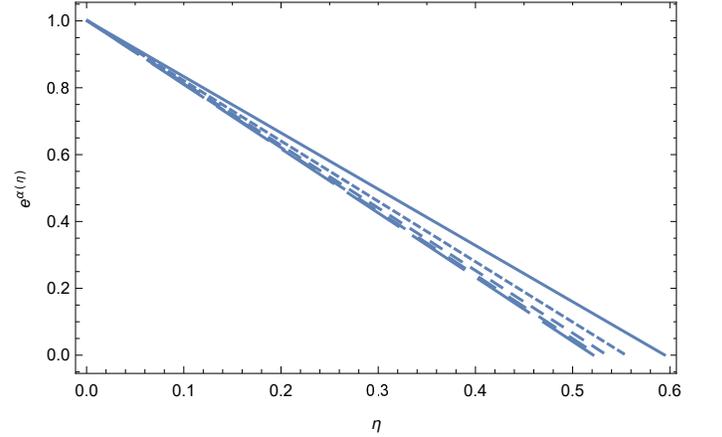}
\caption{Variation of the metric tensor coefficient $e^{\alpha}$ as a function of the  coordinate  $\eta$ for the case of the exponential three-form potential $V=V_0e^{\lambda \zeta}$, for $V_0 =9.9\times 10^{-10}$,  and for different values of $\lambda $: $\lambda=-40$ (solid curve), $\lambda=-120$ (dotted curve), $\lambda =-200$ (short dashed curve), $\lambda=-280$ (dashed curve), and $\lambda =-360$ (long dashed curve), respectively.
} \label{fig13}
\end{figure}

\begin{figure}[htbp]
\centering
\includegraphics[scale=0.7]{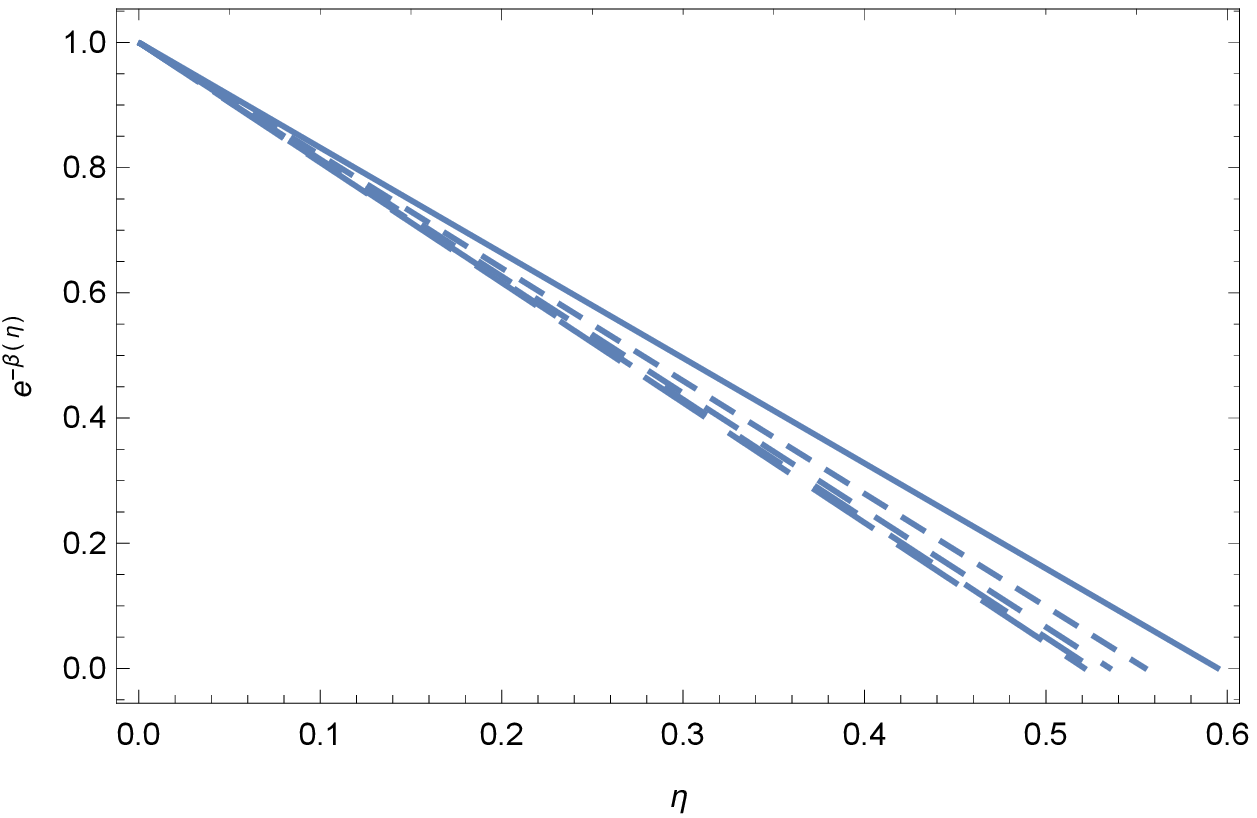}
\caption{Variation of the metric tensor coefficient $e^{-\beta}$ as a function of the coordinate $\eta$  for the case of the exponential three-form potential $V=V_0e^{\lambda \zeta}$, for $V_0 =9.9\times 10^{-10}$,  and for different values of $\lambda $: $\lambda=-40$ (solid curve), $\lambda=-120$ (dotted curve), $\lambda =-200$ (short dashed curve), $\lambda=-280$ (dashed curve), and $\lambda =-360$ (long dashed curve), respectively.
} \label{fig14}
\end{figure}

\begin{figure}[htbp]
\centering
\includegraphics[scale=0.7]{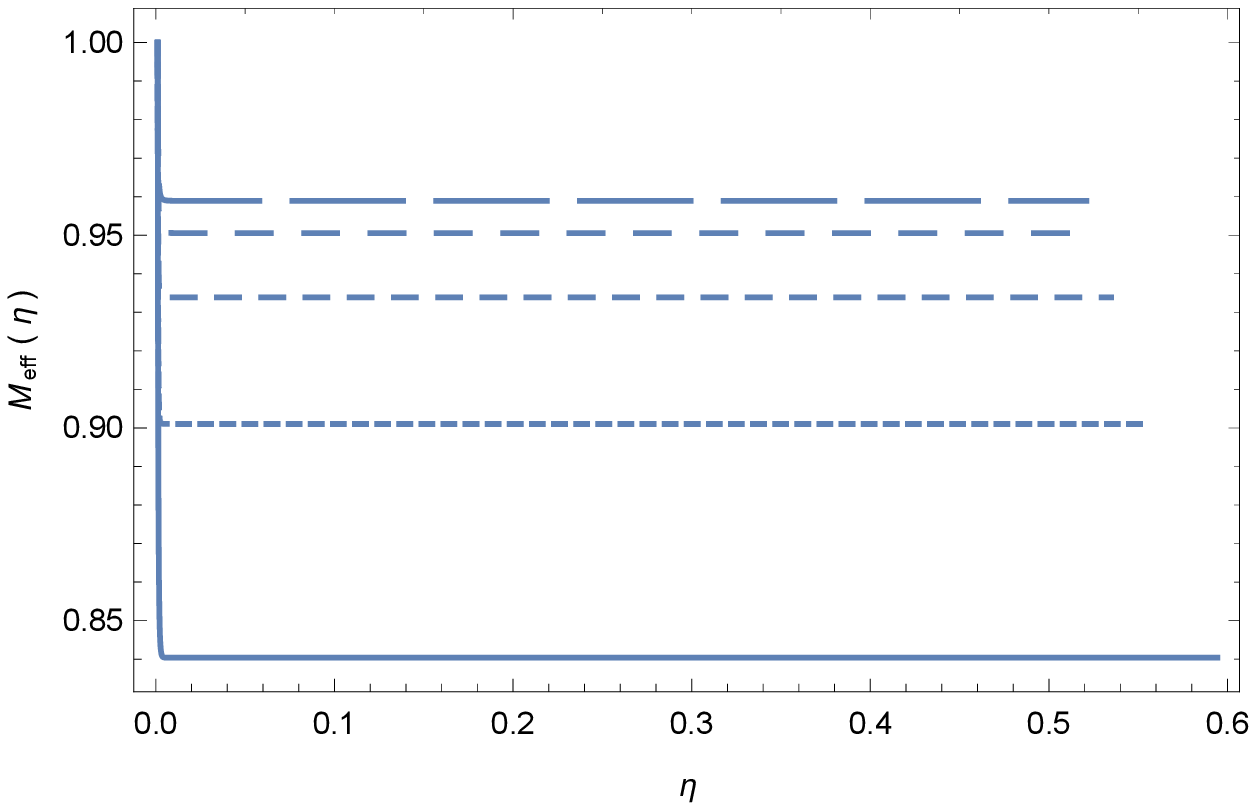}
\caption{Variation of the effective mass $M_{{\rm eff}}$ as a function of the coordinate $\eta$ for the case of the exponential three-form potential $V=V_0e^{\lambda \zeta}$, for $V_0 =9.9\times 10^{-10}$,  and for different values of $\lambda $: $\lambda=-40$ (solid curve), $\lambda=-120$ (dotted curve), $\lambda =-200$ (short dashed curve), $\lambda=-280$ (dashed curve), and $\lambda =-360$ (long dashed curve), respectively.
} \label{fig15}
\end{figure}

\begin{figure}[htbp]
\centering
\includegraphics[scale=0.7]{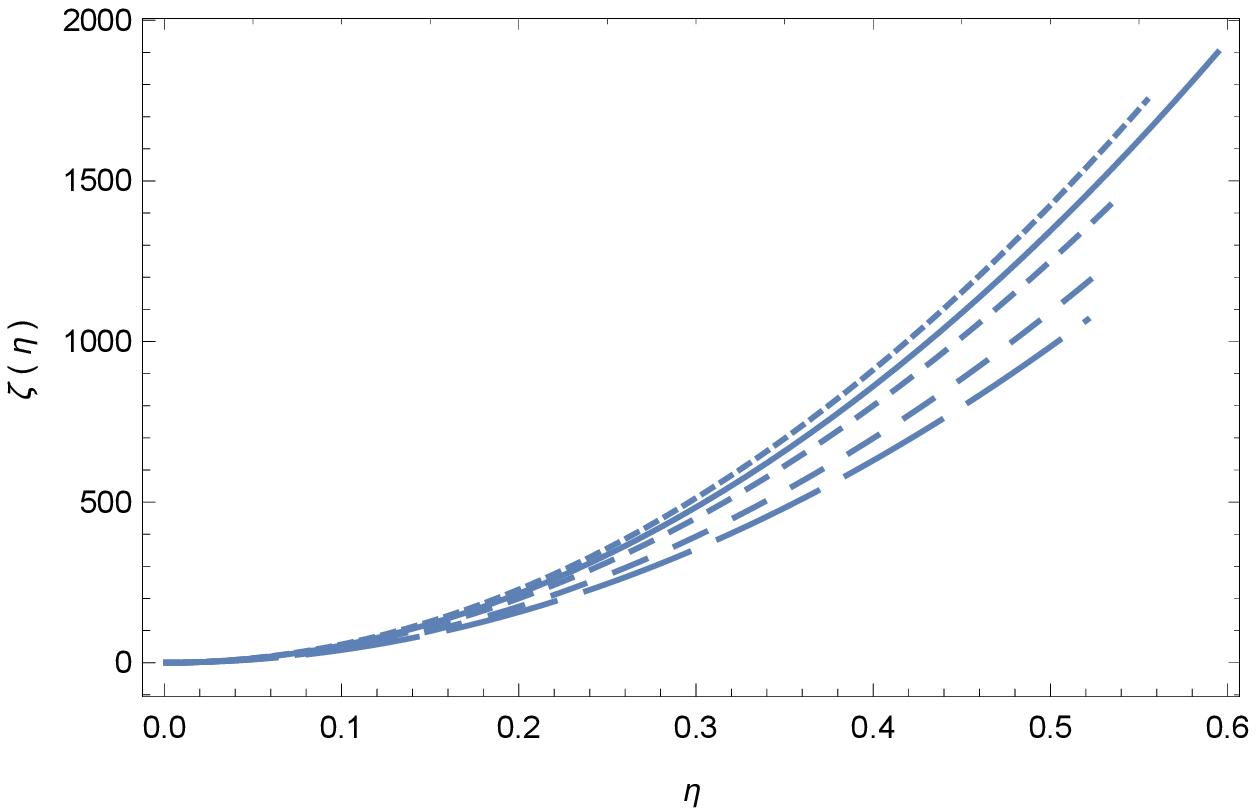}
\caption{Variation of the radial scalar function  $\zeta$ as a function of the coordinate $\eta$  for the case of the exponential three-form potential $V=V_0e^{\lambda \zeta}$, for $V_0 =9.9\times 10^{-10}$,  and for different values of $\lambda $: $\lambda=-40$ (solid curve), $\lambda=-120$ (dotted curve), $\lambda =-200$ (short dashed curve), $\lambda=-280$ (dashed curve), and $\lambda =-360$ (long dashed curve), respectively.
} \label{fig16}
\end{figure}

The static spherically symmetric vacuum gravitational field equations in the presence of a three-form field also admit black hole type solutions. In Figs.~\ref{fig13}-\ref{fig16} we present the results of the numerical integration of the gravitational field equations for $M(0)=1$, $\alpha (0)=0$, $\zeta (0)=10^{-2}$, $\zeta '(0)=10^{-1}$, $V_0=9.9\times 10^{-10}$, and different values of $\lambda$.

As one can see from Figs.~\ref{fig13} and \ref{fig14}, the metric tensor coefficients $e^{\alpha}$ and $e^{-\beta}$ decrease monotonically from their Minkowskian values at infinity to zero, a value reached for a finite value of $\eta=\eta _S$. Hence the compact object possesses an event horizon, and is thus a black hole. The effective mass $M_{{\rm eff}}$, depicted in Fig.~\ref{fig15}, decreases very quickly from its initial value at infinity, and becomes a constant, having the same numerical value from infinity to the event horizon of the black hole. Hence the metric is of the Schwarzschild type, with
\be
e^{\alpha (r)}=e^{-\beta (r)}=1-\frac{2M_{{\rm eff}}\left(V_0,\lambda, \zeta (0),\zeta '(0)\right)}{r},
\ee
with the effective mass, and the position of the event horizon depending on the initial conditions at infinity, and on the parameters of the exponential potential. The radial scalar function increases rapidly from infinity when approaching the event horizon, and takes a finite value for $\eta =\eta _S$.

We can obtain an interpolating  expression for the effective mass function $M_{{\rm eff}}$ in the case of the exponential potential as
\begin{equation}\label{mass_exp}
M_{{\rm eff}}(\eta)\approx A_{MH}^{(\rm exp)},
\end{equation}
with the coefficient $A_{MH}^{(\rm exp)}$  given, for fixed $V_0$ and $\lambda$, by
\bea
A_{MH}^{(\rm exp)}&\approx &0.74 + 9.78 \zeta_0 + 15.812 \frac{\sqrt{\zeta_0}}{u_0} - 146.854 \frac{\zeta_0}{u_0} \nonumber\\
&&+ 0.00134 u_0 - 0.0000528 u_0^2,
\eea
with $R^2=0.997$.

\begin{figure}[htbp]
\centering
\includegraphics[scale=0.7]{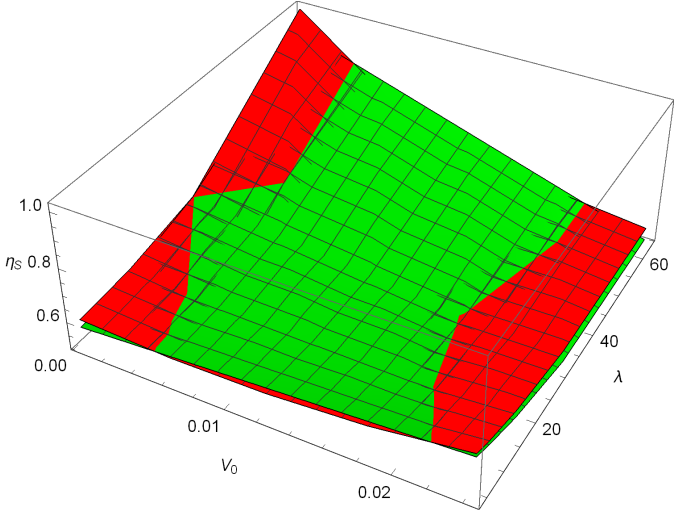}
\caption{Variation of the position of the event horizons of the three-form field supported black holes as a function of the parameters $V_0$ and $\lambda $ of the exponential potential for $\zeta_0=\{0.0004, 0.0064, 0.0124, 0.0184, 0.0244\}$ and $u_0=\{4,8,16,32,64\}$.
} \label{fig16a}
\end{figure}

The distribution of the position of the black holes event horizon supported by a three-form field with exponential potential are represented, for fixed $V_0$ and $\lambda$, in Fig.~\ref{fig16a}.

\section{Thermodynamic properties of black holes}\label{sect5}

In the present Section we consider the thermodynamic properties of the black hole solutions supported by a three-form field. In particular we will concentrate on the surface gravity of the black holes, their Hawking temperature, as well as on their specific heat, entropy and Hawking luminosity. In our investigation of the vacuum field equations in the three-form fields model we have adopted the simplifying assumption that the effective mass function and the lapse function $e^{\alpha}$ are functions of the radial coordinate $r$ only. Hence  the spacetime is static and a timelike Killing vector $t^{\mu}$ exists \cite{Wald,Pielahn:2011ra}. In the present Section, for the sake of clarity, we will restore the physical units in all mathematical expressions.

\subsection{Brief summary of black hole thermodynamics}

For a static black hole that possesses a Killing horizon the definition of the surface gravity $\tilde{\kappa}$  is given by \cite{Wald,Pielahn:2011ra}
\be
t^{\mu}\nabla _{\mu}t^{\nu}=t^{\nu}\tilde{\kappa},
\ee
where $t^{\mu}$ is a Killing vector, and $\nabla _{\mu}$ denotes the covariant derivative with respect to the metric. For a static, spherically symmetric geometry, with the line element given by
\be
ds^2=-\tilde{\sigma} ^2 (r)f(r)c^2dt^2+\frac{dr^2}{f(r)}+r^2d\Omega ^2,
\ee
 we can adopt a suitable normalized Killing vector defined as $t^{\mu}=\left(1/\tilde{\sigma}_{\infty},0,0,0\right)$. Then the surface gravity of the black hole is obtained as \cite{Pielahn:2011ra}
\be
\tilde{\kappa}=\left(\frac{\tilde{\sigma} _{\rm hor}}{\tilde{\sigma} _{\infty}}\right)\frac{c^4}{4GM_{\rm hor}}\left.\left[1-\frac{2GM'(r)}{c^2}\right]\right|_{\rm hor}.
\ee
The subscript hor requires that all physical quantities must be  evaluated on the outer
apparent horizon.  If the function $\tilde{\sigma }\equiv 1$, and $M={\rm constant}$, from the above definition we reobtain the standard result of the surface gravity of a Schwarzschild black hole, which is given by \cite{Wald},
\be
\tilde{\kappa}=\frac{c^4}{4GM_{\rm hor}}.
 \ee

 The temperature $T_{BH}$ of the black hole is obtained as
\be
T_{BH}=\frac{\hbar}{2\pi ck_B} \tilde{\kappa},
\ee
where $k_B$ is Boltzmann's constant. Equivalently, in the variable $r=r_g/\eta$ we obtain for the Hawking temperature of the black hole the expression
\bea
T_{BH}&=&\frac{T_H}{M_{{\rm eff}}\left(\eta _S\right)}\left.\left(1+\eta ^2\frac{dM_{{\rm eff}}\left(\eta\right)}{d\eta }\right)\right|_{\eta =\eta _S}
	\nonumber\\
&=&T_H\left.\theta (\eta)\right|_{\eta =\eta _S},
\eea
where
\be
T_H=\frac{\hbar c^3}{8\pi Gk_BM_{0}},
\ee
 $M_0$ is the standard general relativistic mass of the black hole, and we have denoted
\be
\theta (\eta)=\frac{1}{M_{{\rm eff}}\left(\eta\right)}\left(1+\eta ^2\frac{dM_{{\rm eff}}\left(\eta\right)}{d\eta }\right).
\ee

The specific heat $C_{BH}$ of the black hole is defined as
\bea
C_{BH}&=&\frac{dM}{dT_{BH}}=\left.\frac{dM}{dr}\frac{dr}{dT_{BH}}\right|_{r=r_{\rm hor}},
\eea
and it takes the dimensionless form
\be
C_{BH}=\frac{M_0}{T_H} \left.\frac{dM_{{\rm eff}}\left(\eta \right)}{d\eta}\frac{d\eta}{d\theta }\right|_{\eta =\eta _S}.
\ee

The Hawking entropy $S_{BH}$ of the black hole is obtained as
\bea
S_{BH}&=&\int_{r_{in}}^{r_{\rm hor}}{\frac{dM}{T_{BH}}}=\int_{r_{in}}^{r_{\rm hor}}{\frac{1}{T_{BH}}\frac{dM}{dr}dr},
\eea
or, in a dimensionless form, as
\bea\label{90}
S_{BH}\left(\eta _S\right)=C_H\int_0^{\eta _S}{\frac{1}{\theta \left(\eta \right)}\frac{dM_{{\rm eff}}\left(\eta\right)}{d\eta}d\eta}.
\eea

The black hole luminosity due to the Hawking evaporation can be obtained as
\be
L_{BH}=-\frac{dM}{dt}=-\sigma A_{BH}T_{BH}^4,
\ee
\\
where $\sigma $ is a parameter depending on the adopted physical model, while
\be
A_{BH}=4\pi r_{\rm hor}^2,
\ee
is the surface area of the event horizon. Then for the black hole evaporation time $\tau $ we find
\bea
\hspace{-0.8cm}\tau &=&\int_{t_{in}}^{t_{fin}}{dt}=-\frac{1}{4\pi \sigma}\int_{t_{in}}^{t_{fin}}{\frac{dM}{r_{\rm hor}^2T_{BH}^4}},
\eea
\\
Hence for the black hole evaporation time $\tau $ we obtain the expression
\bea
\hspace{-0.8cm}\tau &=&\int_{t_{in}}^{t_{fin}}{dt}=-\frac{1}{4\pi \sigma}\int_{t_{in}}^{t_{fin}}{\frac{dM}{r_{\rm hor}^2T_{BH}^4}},
\eea
or, in an equivalent dimensionless form,
\bea\label{95}
\tau \left(\eta _S\right)=-\tau _H\int_0^{\eta _S}{\frac{1}{\eta ^2\theta ^4\left(\eta\right)}\frac{dM_{{\rm eff}}\left(\eta\right)}{d\eta }d\eta},
\eea
where we have denoted
\be
\tau _H=\frac{c^4}{8\pi G^2\sigma M_{0}T_{BH}^4}.
\ee

\subsection{Thermodynamics of the Higgs type black holes}

With the help of the interpolating function for the effective mass the Hawking temperature of a black hole supported by a three-form field is obtained explicitly as
\bea
T_{BH}\left(\eta _S\right)&=&T_H\left.\frac{1+\eta ^2\left(B_{MH}+2C_{MH}\eta\right)}{A_{MH}+B_{MH}\eta +C_{MH}\eta ^2}\right|_{\eta =\eta _S}
\eea

\begin{figure}[htbp]
\centering
\includegraphics[scale=0.7]{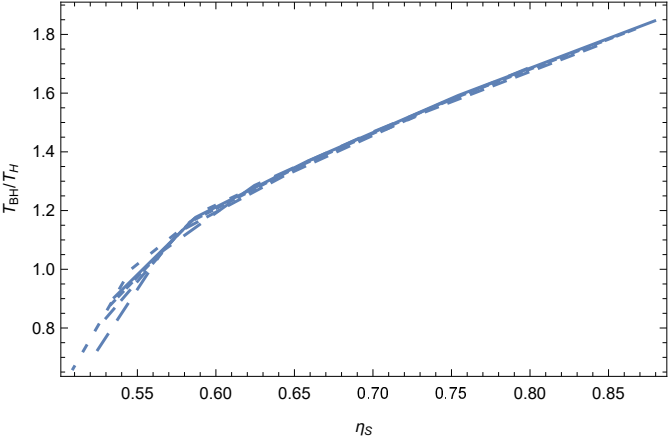}
\caption{Variation of the dimensionless Hawking temperature of the three-form field supported black holes as a function of the event horizon radius  $\eta_S$  for the case of the Higgs type three-form field potential, for $\mu^2=10^{-4}$, $\nu=10^{-2}$, $u_0 \in [40,60]$, and  for different values of $\zeta _0 $: $\zeta _0=10^{-5}$ (solid curve), $\zeta _0=2\times 10^{-5}$ (dotted curve), $\zeta _0 =4\times 10^{-5}$ (short dashed curve), $\zeta _0=8\times 10^{-5}$ (dashed curve), and $\zeta _0 =16\times 10^{-5}$ (long dashed curve), respectively.
} \label{fig17}
\end{figure}

The variation of the Hawking temperature of the three-form field black holes are represented in Fig.~\ref{fig17}.

The Hawking temperature depends on the position of the event horizon of the three-form black hole, and it monotonically increases in time with $\eta _S$. For $\eta _S=0.80$, $T_{BH}\approx 1.6T_H$, while for $\eta _S=0.60$, $T_{BH}\approx 1.2T_H$. The standard Hawking temperature is of the order of $T_H=6.169\times 10^{-8}\times \left(M_{\odot}/M_0\right)$, and for astrophysical type three-form field black holes, having large masses, the shifts in the position of the event horizon produce negligible effects.

\begin{figure}[htbp]
\centering
\includegraphics[scale=0.7]{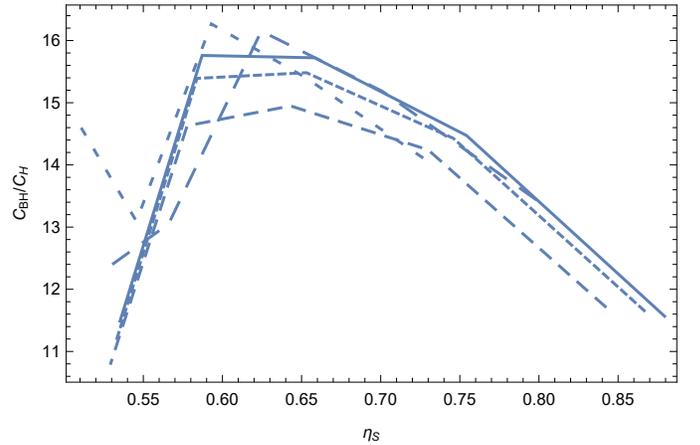}
\caption{Variation of the dimensionless specific heat of the three-form field supported black holes as a function of the event horizon radius  $\eta_S$  for the case of the Higgs type three-form field potential, for $\mu^2=10^{-4}$, $\nu=10^{-2}$, $u_0 \in [40,60]$, and  for different values of $\zeta _0 $: $\zeta _0=10^{-5}$ (solid curve), $\zeta _0=2\times 10^{-5}$ (dotted curve), $\zeta _0 =4\times 10^{-5}$ (short dashed curve), $\zeta _0=8\times 10^{-5}$ (dashed curve), and $\zeta _0 =16\times 10^{-5}$ (long dashed curve), respectively.
} \label{fig18}
\end{figure}

The specific heat of the Higgs type three-form field black hole can be obtained as
\bea
C_{BH}\left(\eta _S\right)&=&\Big\{C_H\left(B_{MH}+2 C_{MH} \eta \right) \times
	\nonumber \\
&&\left[A_{MH}	+\eta  \left(B_{MH}+C_{MH} \eta   \right)\right]^2 \Big\}\Big/
	\nonumber \\
&& \qquad   \Big\{B_{MH} \left(2 A_{MH} \eta +4 C_{MH} \eta ^3 -1\right)
	\nonumber \\
&& \qquad  +\eta  \left[C_{MH} \left(6 A_{MH}
   \eta -2\right)+2
   C_{MH}^2 \eta ^3\right]
	\nonumber \\
 && \qquad +B_{MH}^2 \eta ^2  \Big\},
\eea
%
and its variation with respect to the event horizon is represented in Fig.~\ref{fig18}.

The specific heat of the the three-form black holes has a complicated behavior. After initially increasing as a function of $\eta _S$, $C_{BH}$ reaches a maximum, and then it monotonically decreases towards a minimum value reached at $\eta )S\approx 0.90$. In the range $\eta _S\in (0.50,0.65)$ there is an increase in the numerical values of $C_{BH}$ as compared to the standard general relativistic case, so that for $\eta _S=0.65$, $C_{BH}\approx 1.45C_H$.

\begin{figure}[htbp]
\centering
\includegraphics[scale=0.7]{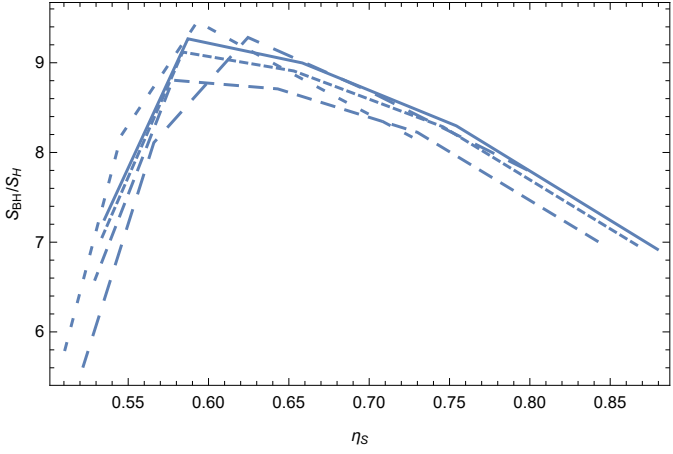}
\caption{Variation of the dimensionless Hawking entropy  of the three-form field supported black holes as a function of the event horizon radius  $\eta_S$  for the case of the Higgs type three-form field potential, for $\mu^2=10^{-4}$, $\nu=10^{-2}$, $u_0 \in [40,60]$, and  for different values of $\zeta _0 $: $\zeta _0=10^{-5}$ (solid curve), $\zeta _0=2\times 10^{-5}$ (dotted curve), $\zeta _0 =4\times 10^{-5}$ (short dashed curve), $\zeta _0=8\times 10^{-5}$ (dashed curve), and $\zeta _0 =16\times 10^{-5}$ (long dashed curve), respectively.
} \label{fig19}
\end{figure}

The variation of the black hole entropy as a function of the event horizon, as given by Eq.~(\ref{90}),  is represented in Fig.~\ref{fig19}.
The behavior of the Hawking entropy of the three-form field black holes has a similar behavior like their specific heat. The entropies are monotonically increasing functions for small $\eta _S$, they reach a maximum, and they decrease for larger values of $\eta _S$. The maximum values of the entropy are of the order $S_{BH}\approx 1.5S_H$.

\begin{figure}[htbp]
\centering
\includegraphics[scale=0.7]{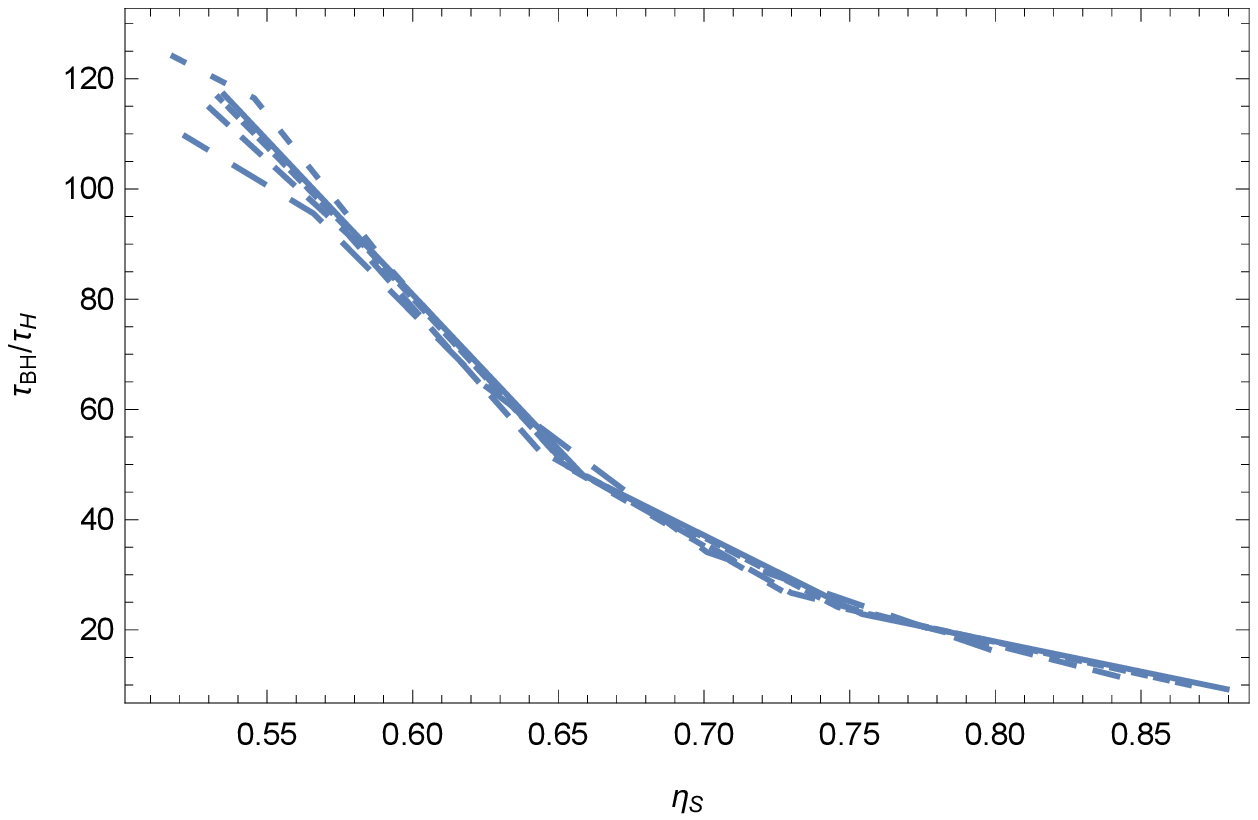}
\caption{Variation of the dimensionless evaporation time of the three-form field supported black holes as a function of the event horizon radius  $\eta_S$  for the case of the Higgs type three-form field potential, for $\mu^2=10^{-4}$, $\nu=10^{-2}$, $u_0 \exists [40,60]$, and  for different values of $\zeta _0 $: $\zeta _0=10^{-5}$ (solid curve), $\zeta _0=2\times 10^{-5}$ (dotted curve), $\zeta _0 =4\times 10^{-5}$ (short dashed curve), $\zeta _0=8\times 10^{-5}$ (dashed curve), and $\zeta _0 =16\times 10^{-5}$ (long dashed curve), respectively.
} \label{fig20}
\end{figure}

With the use of the general Eq.~(\ref{95}),  the ratio of the black hole evaporation time and of the Hawking evaporation time is represented in Fig.~\ref{fig20}. As a function of $\eta _S$ the evaporation time monotonically decreases from a  maximum value $\tau _{BH}\approx 120 \tau _H$, reached for $\eta _S\approx 0.55$, to a value of $\tau _{Bh}\approx 20 \tau _H$ for $\eta _S\approx 0.85$. Since the standard Hawking evaporation time of a black hole is of the order $\tau _H\approx 4.8\times 10^{-27}\times \left(M_0/{\rm g}\right)^3$, it turns out that the evaporation time for three-form field supported black holes can be one or two orders of magnitude higher. Even so, the evaporation time for astrophysical size objects remains very high, with a three-form field black hole having one solar mass completely evaporating via Hawking radiation in around $10^{62}-10^{63}$ years.

\subsection{Exponential potential type black holes}

In the presence of an exponential potential of the three-form field, the metric of the black holes are quasi-Schwarzschild, with the effective mass a constant for most of the range of the variation of $\eta$. Hence the thermodynamical properties of the black holes can be obtained by the thermodynamic properties of the Schwarzschild black holes, with the mass substituted by the effective mass $M_{{\rm eff}}$, as given by Eq.~(\ref{mass_exp}). For the Hawking temperature of the exponential type black hole we obtain
\be
T_{BH}=\frac{\hbar c^3}{8\pi Gk_BM_{{\rm eff}}\left(\eta _S\right)}\approx \frac{\hbar c^3}{8\pi Gk_BA_{MH}^{(\rm exp)}\left(V_0,\lambda, \zeta _0,u_0\right)}.
\ee
Hence the Hawking temperature of the black hole is dependent on the parameters of the potential, and of the initial conditions at infinity of the field $\zeta$. For the specific heat of the black hole we obtain
\be
C_{BH}=-\frac{8\pi k_BG}{\hbar G}M_{{\rm eff}}\approx -\frac{8\pi k_BG}{\hbar G}A_{MH}^{(\rm exp)}\left(V_0,\lambda, \zeta _0,u_0\right).
\ee
There is a dependence of the specific heat on the potential parameters, and on the initial values for $\zeta  $. The negative sign indicates that as a black hole loses mass, and hence energy, its temperature increases. For the entropy of the black hole we can write down the standard expression
\bea
S_{BH}&=&\frac{k_B c^3}{4\hbar G}A_{BH}=\frac{\pi k_Bc^3}{\hbar G}r_{\rm hor}^2
	\nonumber\\
&=&\frac{\pi k_B G M_0^2}{\hbar c}\frac{1}{\eta _S^2\left(V_0,\lambda, \zeta _0,u_0\right)}.
\eea

The numerical value of the black hole entropy is determined by the mass of the central compact object, as well as of the initial condition at infinity of the radial scalar function $\zeta$. Finally, for the rate of the mass loss we have the relation
\be
\frac{dM}{dt}=\frac{\hbar c^4}{15360 \pi G^2}\frac{1}{M^2},
\ee
which yields for the lifetime of the black hole the expression
\bea
t&=&\frac{5120\pi G^2M_0^3}{\hbar c^4}M_{{\rm eff}}^3=\frac{5120\pi G^2 M_0^3}{\hbar c^4}M_{{\rm eff}}^3
	\nonumber\\
&=&\frac{5120\pi G^2M_0^3}{\hbar c^4}\left[A_{MH}^{(\rm exp)}\left(V_0,\lambda, \zeta _0,u_0\right)\right]^3.
\eea

The corrections to the black hole lifetime are given by the third power of the function $A_{MH}^{(\rm exp)}\left(V_0,\lambda, \zeta _0,u_0\right)$, determined by the parameters of the potential and the initial conditions at infinity. However, the evaporation time of the black holes is not significantly influenced by the potential parameters, and the initial conditions.

\section{Discussion and final remarks}\label{conclusions}

In the present paper we have investigated the possible existence of massive compact astrophysical objects, described by black hole and naked singularity type geometries,  in the framework of the three-form field gravitational theory, in which the standard Hilbert-Einstein action of general relativity is extended by the addition of the Lagrangian of a three-form field. In order to investigate the gravitational properties of the model  we have considered the simplest case, corresponding to a vacuum static and spherically symmetric geometry. In this case the system of gravitational field equations depend on the scalar radial function $\zeta$, the radial component of the dual vector $B^{\delta}$ of $A_{\alpha \beta \gamma}$, and on the arbitrary potential $V(\zeta)$ of the three-form field.

Even within the simple vacuum spherically symmetric static model the field equations of the theory become extremely complicated. However, the exact solution of the field equations can be obtained in the case of a constant potential. In this case the metric functions can be obtained in a form similar to the Schwarzschild-de Sitter geometry, but with the solution containing two arbitrary integration constants $c_1$ and $c_2$. Similarly to the metric, the radial scalar function also depends on two arbitrary constants. Depending on the choice of these constants, several types of compact geometries can be obtained. We can first reproduce the standard Schwarzschild-de Sitter geometry, in which the three-form field generates a cosmological constant. However, different choices of the constants are also possible, with the solutions corresponding to the Schwarzschild-anti de Sitter geometry, or, more interestingly, to naked singularities having no event horizon, and with the singularity located at the center $r=0$ of the massive object.

 For arbitrary non-constant potentials $V(\zeta)$ in order to obtain solutions of the vacuum field equations one must use numerical methods. In order to investigate numerically the static spherically symmetric Einstein field equations  in the presence of a three-form field we have reformulated them in a dimensionless form, and, moreover, we have introduced as the independent variable $\eta$ the inverse of the radial coordinate $\eta =r_g/r$. In this representation the numerical integration procedure of the field equations is significantly simplified. On the other hand to proceed with the numerical integration we need  to fix the numerical values of the effective mass, of the radial scalar field $\zeta$, and of its derivative $\zeta '$  at infinity. In our present numerical approach we have assumed that at infinity the geometry is asymptotically flat, and therefore the metric tensor components take their Minkowskian values when $r\rightarrow \infty$. But at infinity the values of $\zeta$, and of $\zeta '$ may be arbitrary (but small), and this choice is consistent with the interpretation of the radial scalar field as a cosmological constant.

When integrating numerically the gravitational field equations two possible behaviors are detected.  The presence of a singular behavior at finite $\eta =\eta _S$ in the field equations, or, more precisely, in the variation with the distance of the metric tensor coefficients, is explained as indicating the existence of an event horizon. We detect the presence of the horizon from the conditions $e^{\alpha \left(\eta _S\right)}=0$ and $e^{-\beta \left(\eta _S\right)}=0$, respectively. Consequently, a singularity at finite $\eta $ corresponds to a black hole type massive astrophysical object. The physical mass of the black hole corresponds to the effective mass of the three-form field model, which is obtained as the sum of the standard mass of the black hole plus the energy contribution from the scalar component of the three-form  field. The second situation is related to the location of the singularity at the center of the massive object $r=0$ only, with the corresponding object representing a naked singularity, an object not covered by an event horizon.

We have considered two classes of numerical solutions of the gravitational field equations in the three-form field theory, corresponding to two choices of the three-form field potential $V (\zeta)$. The potentials we have chosen for our investigations are the Higgs type potential, and the exponential type potential, respectively. In the case of the Higgs potential we have restricted our investigations to initial conditions that lead to the formation of an event horizon, and consequently of black holes. The locations of the event horizons are strongly dependent on the values at infinity of the radial scalar field and of its derivative (the initial conditions), indicating the existence of an important and subtle relation between the initial values of the three-form field at infinity, and the black hole properties. The location of the event horizon is also strongly dependent on the parameters $\mu ^2$ and $\nu$ of the potential Higgs type potential, indicating a complex multi-parametric dependence of the position of the event horizon, and of the black hole properties. The event horizon for three-form field models with Higgs potential can be located at distances of the order of $2r_g$ from the massive object center, at distance much higher  than those of the standard Schwarzschild black holes, indicating the formation of more massive black holes than predicted by standard general relativity. In the case of the Higgs potential the numerical results can be fitted well by some simple analytical that depend on the initial conditions at infinity.

In the case of the exponential potential we have identified numerically two distinct classes of solutions, depending on the initial values of $\zeta$ and $\zeta '$ at infinity, and on the potential parameters. These two classes correspond to naked singularity type solutions, with the singularity located at the center, and no event horizon present, and to standard black holes, characterized by the presence of an event horizon. In both cases the numerical solutions for the metric functions can be approximated well by a Schwarzschild type form with $e^{\alpha}=e^{-\beta}=1\pm 2M_{{\rm eff}}/r$, where $M_{{\rm eff}}$ depends on the initial conditions at infinity, and on the parameters of the exponential potential. The positive sign indicates the presence of a naked singularity, while the negative sign corresponds to an object with an event horizon.
The analytical fittings of the numerical results are extremely useful in the study of the thermodynamic properties of the three-form field black holes. They also greatly simplify the study of the dynamics and motion of matter particles around black holes and naked singularities.

One important question in the physics of the naked singularities is if the energy conditions are satisfied for this type of objects. For example, in \cite{Harko:2000ni} it was shown that in the case of a mixture of a null charged strange quark fluid and radiation collapsing in a Vaidya space-time a naked singularity can be formed, with the matter satisfying the weak, strong and dominant energy conditions. The formation of a naked singularity or of a black hole depends  on the initial distribution of the density and velocity,  and on the constitutive nature of the collapsing matter. In the case of the naked singularities generated by the three form fields one could also formulate a number of energy conditions, similar to the case of ordinary matter, as $\rho >0$, $p_r>0$, $p>0$, $\rho >p_r>0$, and $\rho >p>0$. However, from a physical point of view, these conditions refer to some effective quantities constructed from the three form field, and generally they are not satisfied for arbitrary three form field configurations.

The problem of the existence in nature of the naked singularities is closely related to their stability properties. The Reissner-Nordstr\"{o}m naked singularity with $|Q|>M$ is unstable under small linear perturbations, and a similar instability occurs for the rotating (Kerr) solution with angular momentum $a>M$ \cite{Dotti}. These results suggest that these spacetimes cannot be the endpoint of physical gravitational collapse. In \cite{Christ} it was shown that as a result of the gravitational collapse of a spherically symmetric scalar field the formation of the naked singularities occurs, but this phenomenon is unstable. However, as compared to the above mentioned types of naked singularities, those formed in the presence of a three form field are of a modified Schwarzschild type, with the sign of the term $M/r$ changed from negative to positive. The problem of their stability/instability is a very interesting one, due to the different nature of the metric. It is known that the Schwarzschild black hole is linearly stable under gravitational and electromagnetic perturbations \cite{Dotti1}, so that one may conjecture that a similar property holds for a three form field supported by Schwarzschild type naked singularities. Hence the investigation of the stability properties of the three form field naked singularities may lead to different results as compared to the Reissner-Nordstr\"{o}m, Kerr or scalar field collapse cases.

We have also investigated in detail the thermodynamic properties of the three-form field black hole solutions obtained by numerical methods. The Hawking temperature is an interesting and essential physical property of black holes. The horizon temperature of the three-form black holes indicates a strong dependence on the initial conditions at infinity of the radial scalar field, and of the properties of the radial scalar field potential. This is very different from the properties of the standard general relativistic Hawking temperature, which depends only on the  mass of the black hole, and is independent of the asymptotic conditions at infinity. Similar properties characterize the behavior of the specific heat, entropy and evaporation time of the three-form field black holes. The numerical results show that these quantities are also strongly dependent on the initial conditions of the radial scalar field $\zeta$ at
infinity.

Moreover, in the three-form field gravitational theory the black hole evaporation times may be very different as compared to the similar results in standard general relativity. But we should note that our results on the thermodynamics of black holes, obtained for only two radial scalar field potentials and for a limited range of initial conditions at infinity may be considered of qualitative nature only. But even at this qualitative level they show the complexity of the compact objects supported by the three-form fields, and of the interesting physical and astrophysical processes related to them.
In particular, the analytic representations of the numerical results may be applied for the study of the electromagnetic properties of the thin accretion disks existing around black holes. These properties may help in discriminating the three-form field black holes from other  similar theoretical objects as well as from their general relativistic counterparts, and for allowing to obtain observational constraints on the model parameters.

The no-hair theorem \cite{Bekenstein:1971hc, Bekenstein:1972ky, Adler:1978dp, Bek2} is an important result in black hole physics. It states that asymptotically flat black holes do not allow for the presence of external nontrivial scalar fields, with non-negative field potential $V (\phi)$. The results obtained in our present investigations indicate that in its standard formulation the no-hair theorem  cannot be extended to the static spherically symmetric solutions of the three-form field gravitational theory. All the numerical black hole solutions we have obtained are asymptotically flat, and three-form fields with positive radial scalar field potentials exist around them. Similar results have been obtained for the black hole solutions in the the hybrid metric-Palatini gravity theory \cite{Danila:2018xya}, suggesting that the no-hair theorems may not be valid in some modified gravity theories.
But the answer to the question if such properties are a result of the particular choice of the three-form field radial scalar potentials, of the initial conditions and of the scalar potential parameters, or that they are some generic properties of the theory, requires  further and detailed investigations, at both theoretical and computational levels.

Different types of physical fields may play an important role in astrophysics and cosmology, especially as potential constituents of the dark energy and dark matter components of the Universe. In particular the role of the scalar fields has been intensively investigated. From a qualitative point of view, since there is a dual representation of the 3-form theory to a scalar field \cite{Mulryne:2012ax,Germani:2009iq} one may explore particular astrophysical settings describing objects such as oscillatons \cite{Alcubierre:2003sx,Seidel:1991zh} or maybe, their complex relatives, boson stars \cite{RevBS1,RevBS2}, in the scalar representative frame. In the latter case, since boson stars are usually constructed from a complex scalar field (decomposed as two real scalar fields), the existence of the dual description from a single 3-form field could be arguable. However, it is important to note that this dual nature, 3-form $\leftrightarrow$ scalar field, breaks down in some cases, even for fairly simple self interactions \cite{Koivisto:2009fb}, depending on the choice for the potential $V(A^2)$. This fact however, is not problematic, and, on the contrary, simply suggests that 3-forms can provide us with new physics upon richer cosmological and astrophysical settings, undoubtedly worth exploring.

In particular, one may mention Bose-Einstein condensates consisting of ultralight bosons, and which can form localized and coherently oscillating stellar type configurations \cite{HaCh}. For bosons having a higher mass, the bounded configurations may have typical masses and sizes of the order of magnitude similar to those of the neutron stars. Such objects formed from a primordial scalar field are known as boson stars (for reviews on the structure and properties of boson stars see \cite{RevBS1} and \cite{RevBS2}, respectively). Usually boson stars are constructed from a complex scalar field coupled to gravity, with the scalar field $\phi \left(t,\vec{r}\right)$ decomposed into two real scalar fields $\phi_R\left(t,\vec{r}\right)$ and $\phi_I\left(t,\vec{r}\right)$ so that $\phi \left(t,\vec{r}\right)=\phi_R\left(t,\vec{r}\right)+i\phi_I\left(t,\vec{r}\right)$ \cite{RevBS2}. The evolution and properties of a boson star can be obtained from the action \cite{RevBS2}
\be
S_{BS}=\int d^4 x \sqrt{-g}\left( \frac{1}{2\kappa^2}R+\mathcal{L}_{SF} \right),
\ee
where
\be
\mathcal{L}_{SF} = - \frac{1}{2} \left[ g^{\mu \nu} \nabla _{\mu} \bar{\phi }\, \nabla _{\nu} \phi + V\left( \left| \phi \right| ^2\right) \right] ,
\ee
where $\bar{\phi }$ is the complex conjugate of the field, while $V(|\phi |^2)$ is the scalar field potential depending only on the magnitude of the scalar field. There is a formal analogy between scalar field models, and the three form field model investigated in the present paper. In both cases the action is constructed from the standard Hilbert-Einstein term plus the field Lagrangian, which in our approach is given by Eq.~(\ref{lagrangian}). Hence if one could map in the complex plane the 3-form Lagrangian $\mathcal{L}_A\mapsto\mathcal{L}_{SF}$, then the present three form field extension of Einstein gravity would become equivalent with a scalar field model, and the corresponding solutions, assuming that the mapping three forms field $\rightarrow$ scalar field does exist,  would describe boson stars of different types.

In our analysis we have considered as a particular case of the three-form field potential the Higgs type potential, given by Eq.~(\ref{Hpot}).  In the case of boson stars the gravitationally bounded configuration with the quartic self-interaction potential $V\left( \left| \phi \right| ^2 \right) = m^2 \left| \phi \right| ^2 \, + \frac{\lambda }{2} \left| \phi \right| ^4$ was investigated in \cite{Colpi}. The physical properties of such a potential can be parameterized by the quantity $\Lambda =\lambda M_{\rm Planck}^2/4\pi G m^2$, where $M_{\rm Planck}$ is the Planck mass. The maximum mass of a boson star with quartic self-interaction potential is given by $ M_{\max } \approx 0.22 \Lambda ^{1/2} M_{\rm Planck}/m $ \cite{RevBS2,Colpi}. It is also interesting to point out that in the Thomas-Fermi limit the scalar field becomes equivalent with a fluid, which in the low or moderate density limit has the equation of state $P\propto \rho ^2$ \cite{Colpi}.

By assuming that the three form field-scalar field correspondence is valid, by parameterizing the Higgs type potential with the help of the parameter $\Lambda =\nu/4\pi G\mu ^2$, the maximum mass of a stable three form field star is given by $M_{\max}\propto \sqrt{\nu}/\sqrt{4\pi G}\mu ^3$, which could lead to masses of the same order of magnitude as the Chandrasekhar limit for neutron stars, $M_{\max}\approx 3M_{\odot}$. However, the similarity (or equivalence) between three form field stars and boson stars strongly relies on the fulfilling of the condition $p_r\propto \rho ^2$, which, with the use of Eqs.~(\ref{T1}) and (\ref{T2}) becomes
\be
F^2-V\propto \left(F^2/48-V+\zeta V_{,\zeta}\right)^2.
\ee

If this differential equation has a solution for $\zeta$, then the corresponding three form field stellar model has similar properties to a boson star. On the other hand, if no such solution for the radial function $\zeta$ exists, then in the absence of ordinary matter the modified Einstein gravity in the presence of the three form fields admits only black hole type solutions.  We would like to point out that the above conclusions about the relations between boson and three form stars are mostly of qualitative nature, and in order to fully clarify these issues a detailed investigation of the properties of the three form field stars is required. From a technical point of view one can identify the formation of a star-like object from the the conditions $e^{\alpha \left(\eta _S\right)}\neq 0$ and $e^{-\beta \left(\eta _S\right)}\neq 0$, respectively. Thus, as mentioned above, since the physical properties of the solutions of the Einstein field equations in the presence of a three form field are strongly dependent on the parameters $\mu ^2$ and $\nu$ of the Higgs potential the formation of stellar type objects with no event horizon is possible.

In summary, the three-form field gravitational theory possess a rich mathematical structure, with theoretical properties that generate an intricate external dynamics. Consequently, the properties of the black holes in the three-form field theory are more complex as compared with the standard general relativistic black holes. These properties are related to the intrinsic properties of the three-form fields, which, from a mathematical point of view,  lead to very complicated, strongly nonlinear, gravitational field equations. The new physical and geometrical effects generated by the presence of the three-form fields  can also lead to some specific astrophysical and cosmological effects, whose observational detection may open new perspectives in the testing of gravitational theories. The  astrophysical and observational implications of the black holes supported by a three-form field will be investigated  in a future publication.

\section*{Acknowledgements}

We would like to thank to the anonymous referee for comments and suggestions that helped us to improve our manuscript. BJB is supported by Funda\c{c}\~ao para a Ci\^encia e a Tecnologia (FCT, Portugal) through the grant PD/BD/128018/2016. BD is partially supported by a grant from the UEFISCDI PNIII-P1-1.2-PCCDI-2017-266 "SAFESPACE" Contract, Nr. 16PCCDI/01.03.2018.
FSNL acknowledges support from the Funda\c{c}\~{a}o para a Ci\^{e}ncia e a Tecnologia (FCT) Scientific Employment Stimulus contract with reference CEECIND/04057/2017.
BJB and FSNL also thank funding from the research grants No. UID/FIS/04434/2019 and No. PTDC/FIS-OUT/29048/2017.




\end{document}